\documentclass{aa}  

\usepackage{graphicx}
\usepackage{natbib}
\usepackage{enumerate}
\usepackage{txfonts}
\begin{document}

   \title{2P/Encke, the Taurid complex NEOs and the Maribo and Sutter's Mill meteorites.
\thanks{Based on observations performed at the European Southern Observatory, Paranal, Chile: Program 087.C-0788(A).}
   }


   \author{C. Tubiana
          \inst{1}
          \and
          C. Snodgrass
           \inst{2,1}
          \and
          R. Michelsen
          \inst{3}
          \and   
          H. Haack
          \inst{3}
          \and
          H. B\"ohnhardt
          \inst{1}
          \and
          A. Fitzsimmons
          \inst{4}
          \and
          I. P. Williams
          \inst{5}
          }

   \institute{Max Planck Institute for Solar System Research, Justus-von-Liebig-Weg 3, 33077 G\"ottingen, Germany\\
              \email{tubiana@mps.mpg.de}
         \and
    Planetary and Space Sciences, Department of Physical Sciences, The Open University, Milton Keynes, MK7 6AA, UK    
     \and
         		Centre for Star and Planet Formation, Natural History Museum of Denmark, University of Copenhagen, Denmark.
	\and
		Astrophysics Research Centre, Department of Physics and Astronomy, Queen's University Belfast, UK
	\and
		School of Physics and Astronomy, Queen Mary, University of London, UK
	}

   \date{Received: 12/12/2014 ; Accepted: 29/09/2015}

 
  \abstract
   {}
   {2P/Encke is a short period comet that was discovered in 1786 and has been extensively observed and studied for more than 200 years.  The Taurid meteoroid stream has long been linked with 2P/Encke owing to a good match of their orbital elements, even though the comet's activity is not strong enough to explain the number of observed meteors. Various small near-Earth objects (NEOs) have been discovered with orbits that can be linked to 2P/Encke and the Taurid meteoroid stream. Maribo and Sutter's Mill are CM type carbonaceous chondrite that fell in Denmark on January 17, 2009 and April 22, 2012, respectively. Their pre-atmospheric orbits  place them in the middle of the Taurid meteoroid stream, which raises the intriguing possibility that comet 2P/Encke could be the parent body of CM chondrites. }
   {To investigate whether a relationship between comet 2P/Encke, the Taurid complex associated NEOs, and CM chondrites exists, we performed photometric and spectroscopic studies of these objects in the visible wavelength range. We observed 2P/Encke and 10 NEOs on August 2, 2011 with the FORS instrument at the 8.2 m Very Large Telescope on Cerro Paranal (Chile).}
   {Images in the R filter, used to investigate the possible presence of cometary activity around the nucleus of 2P/Encke and the NEOs, show that no resolved coma is present. None of the FORS spectra show the 700 nm absorption feature due to hydrated minerals that is seen in the CM chondrite meteorites. All objects show featureless spectra with moderate reddening slopes at $\lambda$ < 800 nm.  Apart for 2003~QC$_{10}$ and 1999~VT$_{25}$, which show a flatter spectrum, the spectral slope of the observed NEOs is compatible with that of 2P/Encke. However, most of the NEOs show evidence of a silicate absorption in lower S/N data at $\lambda$ > 800 nm, which is not seen in 2P/Encke, which suggests that they are not related.}
   {Despite similar orbits, we find no spectroscopic evidence for a link between 2P/Encke, the Taurid complex NEOs and the Maribo and Sutter's Mill meteorites. However, we cannot rule out a connection to the meteorites either, as the spectral differences may be caused by secondary alteration of the surfaces of the NEOs.}


   \maketitle
%

\section{Introduction}
\label{sec-intro}
Comet 2P/Encke is a short-period comet that was discovered in 1786 and it has been studied for more than 200 years. It has an 
orbital period of 3.3 years and its orbit is dynamically decoupled from Jupiter's control because of gravitational interaction with terrestrial planets \citep{Levison2006}. It is the 
only comet known to be on such an orbit (107P/Wilson-Harrington has a similar orbit, but its status as an active comet is questionable, because it does not show repetitive activity each orbit). 
The lack of close encounter with Jupiter results in a very smooth orbital evolution \citep{Levison2006} but raises an interesting problem relating to the origin of this comet, since such encounters are essential if it is to be captured from the outer solar system. A slow orbital evolution onto its present orbit is also unlikely since all the volatiles would have been exhausted. There are two interesting possibilities to explain this: Either comet 2P/Encke has an origin within the main asteroid belt -- recent discoveries \citep{Hsieh2006,Jewitt2012AJ} re-enforce this possibility -- or there was a major fragmentation of an earlier larger comet that led to significant changes in the aphelion distance of the resulting fragments \citep{Whipple1940}. 

The nucleus of 2P/Encke is dark (geometric albedo of 0.047 $\pm$ 0.023 \citep{Fernandez2000ic}), has an effective radius of 2.4 $\pm$ 0.3 km \citep{Fernandez2000ic}, and it has polarimetric properties that are unique compared to other measured types of solar system objects, such as asteroids, TNOs, cometary dust, and Centaurs \citep{Boehnhardt2008aa}. The colours of 2P/Encke's nucleus are typical for comets \citep{Jewitt2002,Snodgrass2006mnras}, and a noisy spectrum reveals a moderate red slope \citep{Luu1990}. 

The Taurid meteoroid stream has long been associated with 2P/Encke because of the similarity in their orbital elements. However the activity of 2P/Encke at the present time is not strong enough to explain the high number of meteoroids in the stream. \citet{Whipple1940} suggested that several thousand years ago a giant comet fragmented that resulted in 2P/Encke and several other bodies within the meteoroid complex, as well as a large number of meteoroids. This general hypothesis is supported by the fact that numerous authors agree (see \citet{Jopek2013MNRAS} for a list) that the stream is, in fact, a complex of several smaller meteoroid streams and filaments. The most recent list of such sub-streams is given by \citet{Porubcan2006}. Further support for this hypothesis came when \citet{Clube1984MNRAS} showed that several Apollo asteroids had orbits that were very similar to the Taurid stream while a decade later, \citet{Asher1993MNRAS} suggested that the complex of meteoroid streams, comet 2P/Encke, and the then known associated Apollo asteroids, could all have been formed by the fragmentation of a giant comet 20 to 30 ky ago.
Since that date several, other authors have claimed associations between NEOs and the Taurid complex, the present number of candidates being well in excess of 100.
Orbital evolution is fairly rapid in this locality of the solar system and most asteroids have low inclination, many of the supposed orbital similarities are fortuitous, with orbits being similar at the present time, but implying no generic connections. \citet{Porubcan2004} argues that, to claim any generic relationships, the orbits of the NEOs and those of the meteoroid streams need to have remained similar for a long time, which they chose to be 5000 years. \citet{Porubcan2006} applied this criterion to the Taurids, resulting in the number of associated NEOs dropping from 100 to around 10. 

All the claimed associations between comet 2P/Encke, the Taurid meteoroid stream, and some NEOs are based on similarities of their orbits. In addition to dynamical properties, common taxonomic properties can also provide an indication of a common origin for small bodies in the solar system. Taxonomic properties are poorly-known for cometary nuclei, and only few comets have spectroscopic measurements in the visible wavelength range. The existing spectra of bare nuclei are generally featureless and display different reddening slopes \citep{Lamy2004comets}. Given the poor S/N ratio that is usually obtained in observations, more subtle features, such as ones from hydrated minerals, are beyond the detection limit in most cases. Multi-colour photometry is available for a few more nuclei and is all consistent with featureless red slopes \citep{Snodgrass2006mnras,Lamy+Toth09}.

If the Taurid complex NEOs are fragments of the same body as 2P/Encke, we expect them to have the same spectral properties as the comet nucleus. Furthermore, it would be reasonable to expect that these NEOs could show cometary activity. Recently, \citet{Popescu2011,Popescu2014} published near infra-red spectra of NEOs, including some associated with the Taurid complex, with the goal of testing to see if proposed members have comet-like spectral types and to compare the resulting spectra with meteorite types. While limited to only the larger proposed members, these studies found that a majority of the NEOs sampled had S-complex spectroscopic classification and were best matched by ordinary chondrite meteorite samples.

Maribo and Sutter's Mill are CM chondrites that fell on January 17, 2009 \citep{Haack2012M} and on April 22, 2012 \citep{Jenniskens2012}. The pre-atmospheric orbits of the two meteorites are very similar and places them right in the middle of the Taurid meteoroid stream \citep{Haack2011, Jenniskens2012}. This raises the intriguing possibility that comet 2P/Encke could be the parent body of CM chondrites, meaning that these meteorites are potentially samples of cometary material we can study in the laboratory. CM chondrites show signs of extensive aqueous alteration, which suggest that the parent body was an icy body that was at least partially molten at some point. It is therefore possible that the parent body of the CM chondrites is a comet \citep{Lodders1999SSRv}. However, \citet{Jenniskens2012} argued that CM chondrites are more likely to come from main belt asteroids since there is no evidence of aqueous alteration in Jupiter family comets. 

To investigate whether a relationship between comet 2P/Encke, the Taurid complex associated NEOs and CM chondrites exists, we performed spectroscopic studies of these objects. By observing in the visible with an 8m class telescope we could sample fainter objects within the Taurid complex, and observe the comet nucleus directly, adding to the work on brighter objects by \citet{Popescu2014}. We also perform a direct search for cometary activity around the NEOs.

\section{Observations and data reduction}
We observed 2P/Encke and 10 NEOs in the visible wavelength range on August 2, 2011 with the FORS instrument at the 8.2 m Very Large Telescope on Cerro Paranal (Chile). The observational circumstances are summarised in Table \ref{obs-table}. FORS \citep{Appenzeller1998} is the visual and near UV Focal Reducer and low-dispersion Spectrograph for the VLT. Its detector consists of a mosaic of two 2k $\times$ 4k MIT CCDs (15 $\mu$m). In the standard resolution mode, which we used for the observations, the image scale is 0.25\arcsec /pix, providing a field of view of 6.8\arcmin $\times$ 6.8\arcmin. For more details about FORS see: http://www.eso.org/sci/facilities/paranal/instruments/fors.html.
The observations were performed with the telescope tracking at the proper motion rate of the objects. \newline

\begin{figure*}
\centering
   \includegraphics[width=0.65\columnwidth,trim=3.5cm 0.5cm 7.5cm 16.5cm,clip=true]{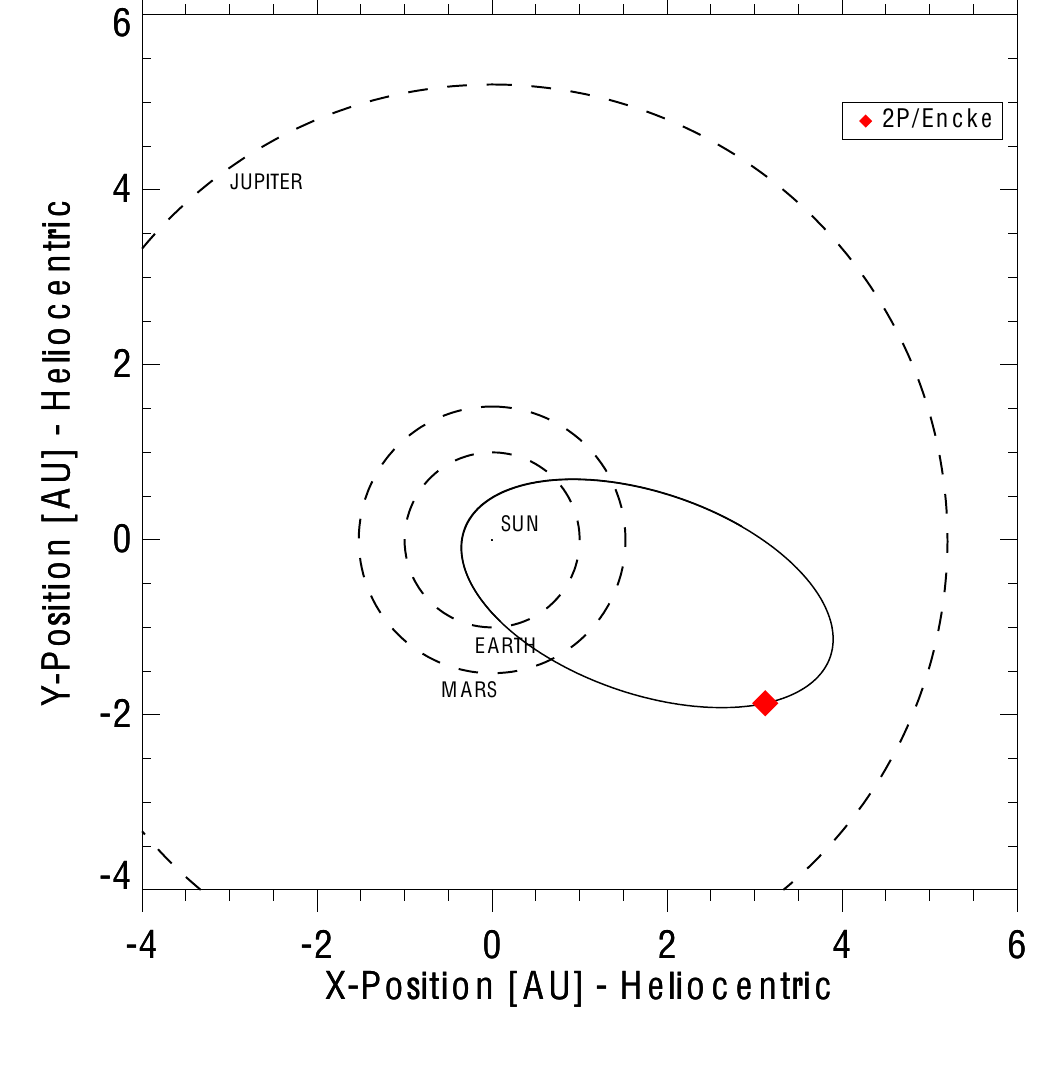}
   \includegraphics[width=0.65\columnwidth,trim=3.5cm 0.5cm 7.5cm 16.5cm,clip=true]{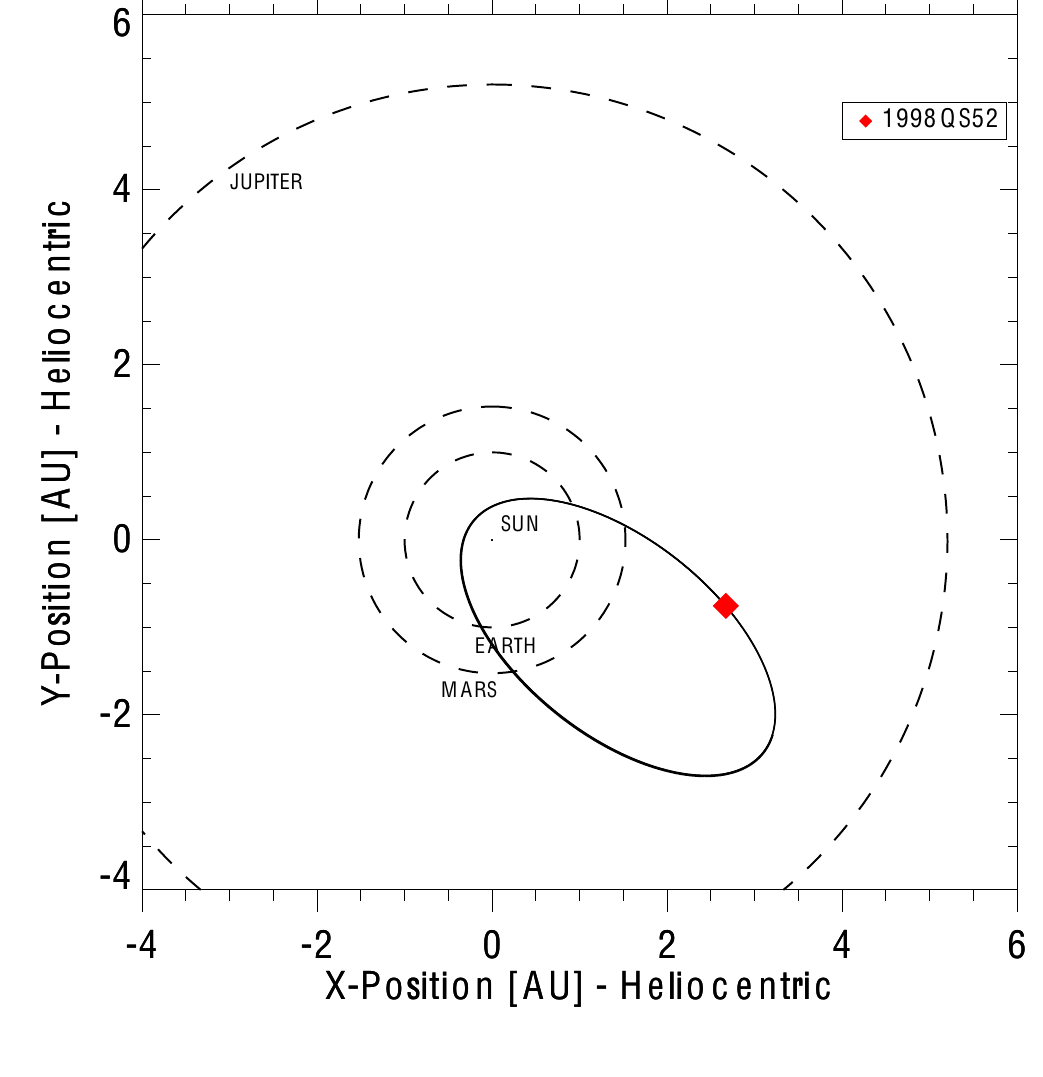}
   \includegraphics[width=0.65\columnwidth,trim=3.5cm 0.5cm 7.5cm 16.5cm,clip=true]{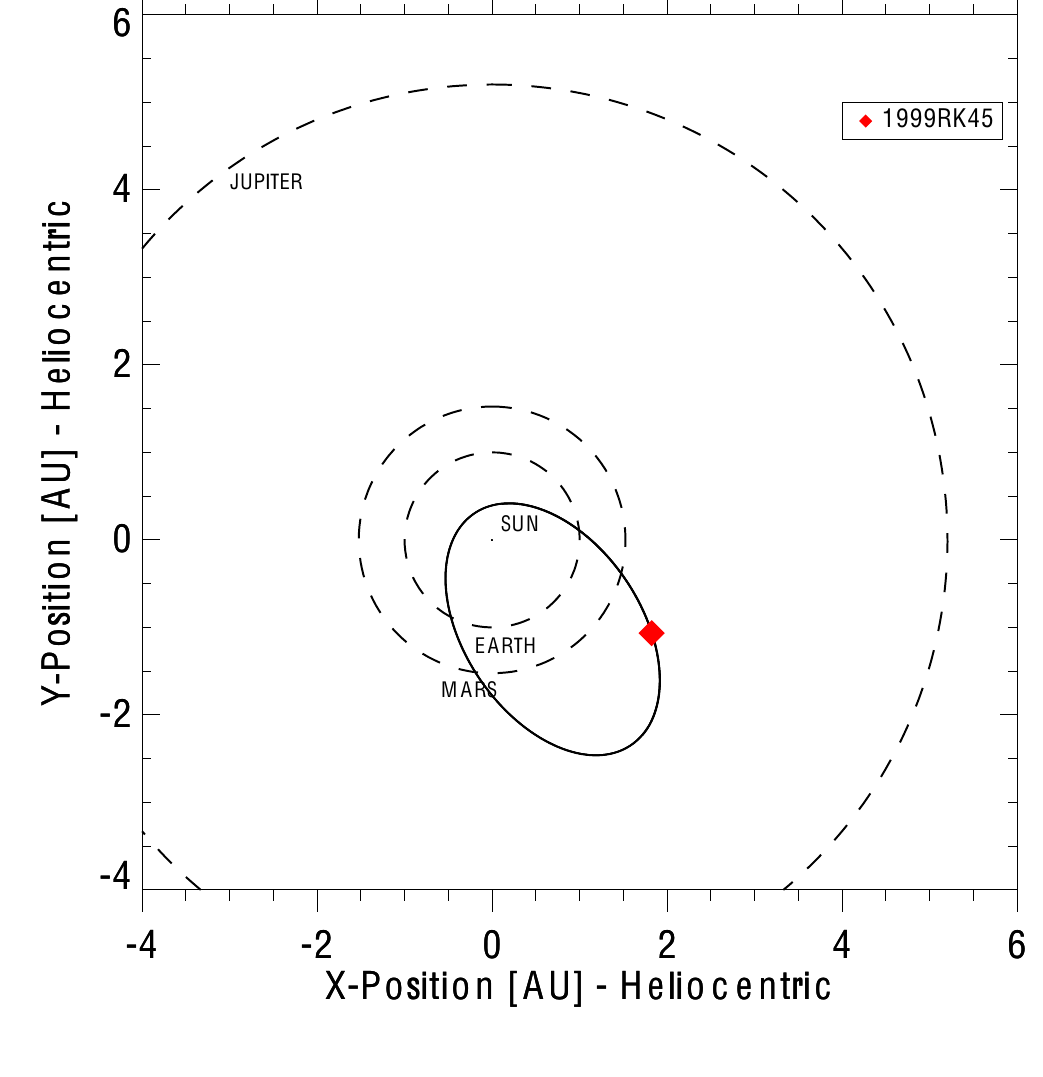}
   \includegraphics[width=0.65\columnwidth,trim=3.5cm 0.5cm 7.5cm 16.5cm,clip=true]{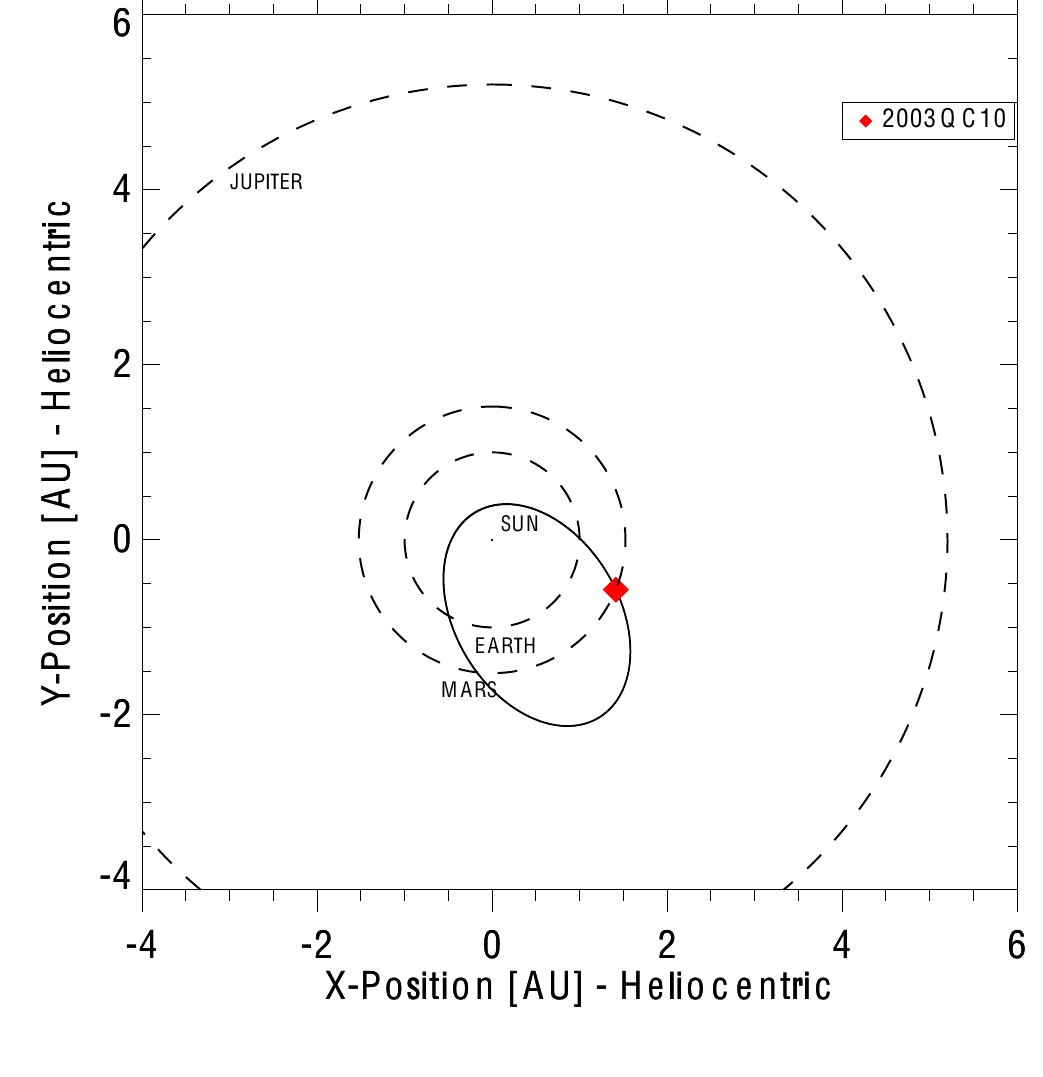}
   \includegraphics[width=0.65\columnwidth,trim=3.5cm 0.5cm 7.5cm 16.5cm,clip=true]{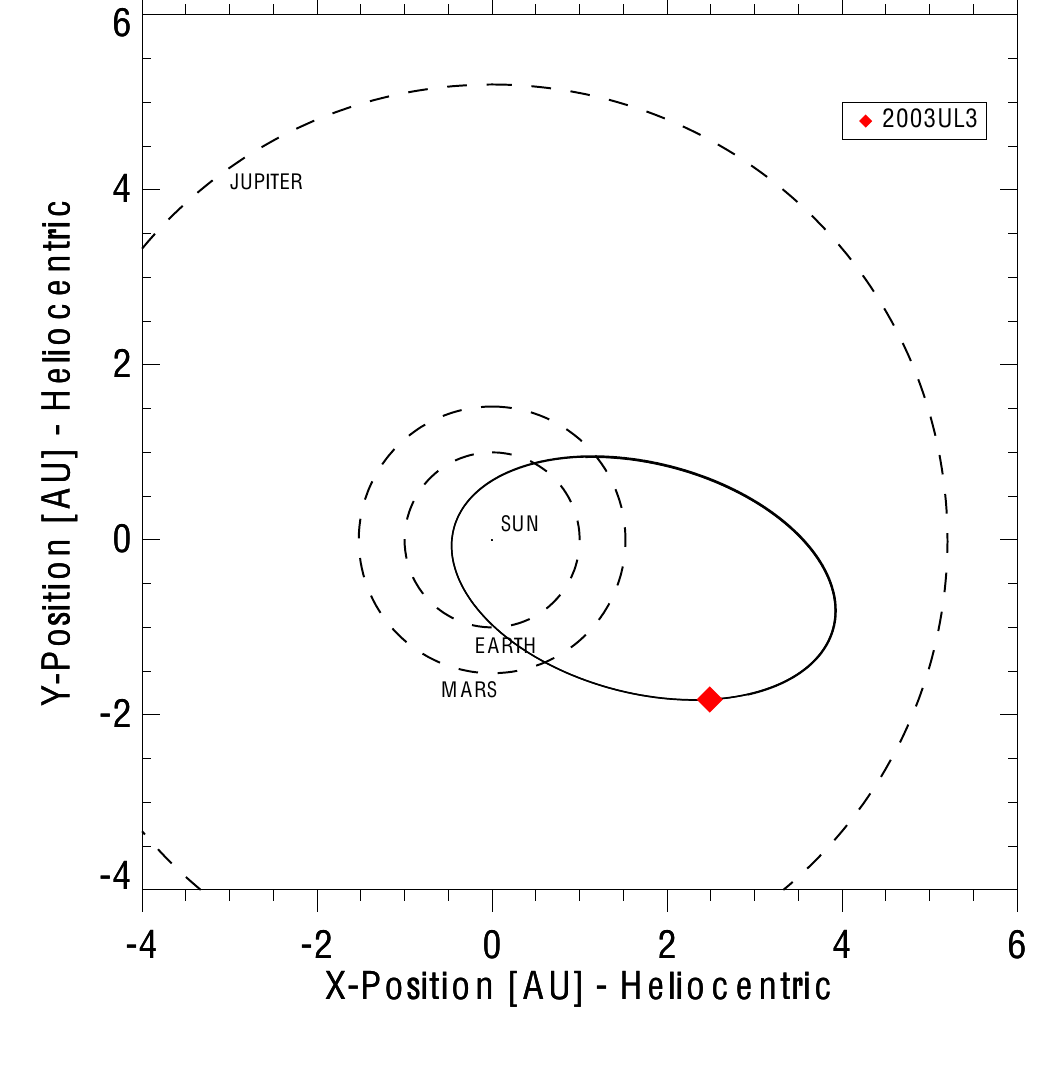}
   \includegraphics[width=0.65\columnwidth,trim=3.5cm 0.5cm 7.5cm 16.5cm,clip=true]{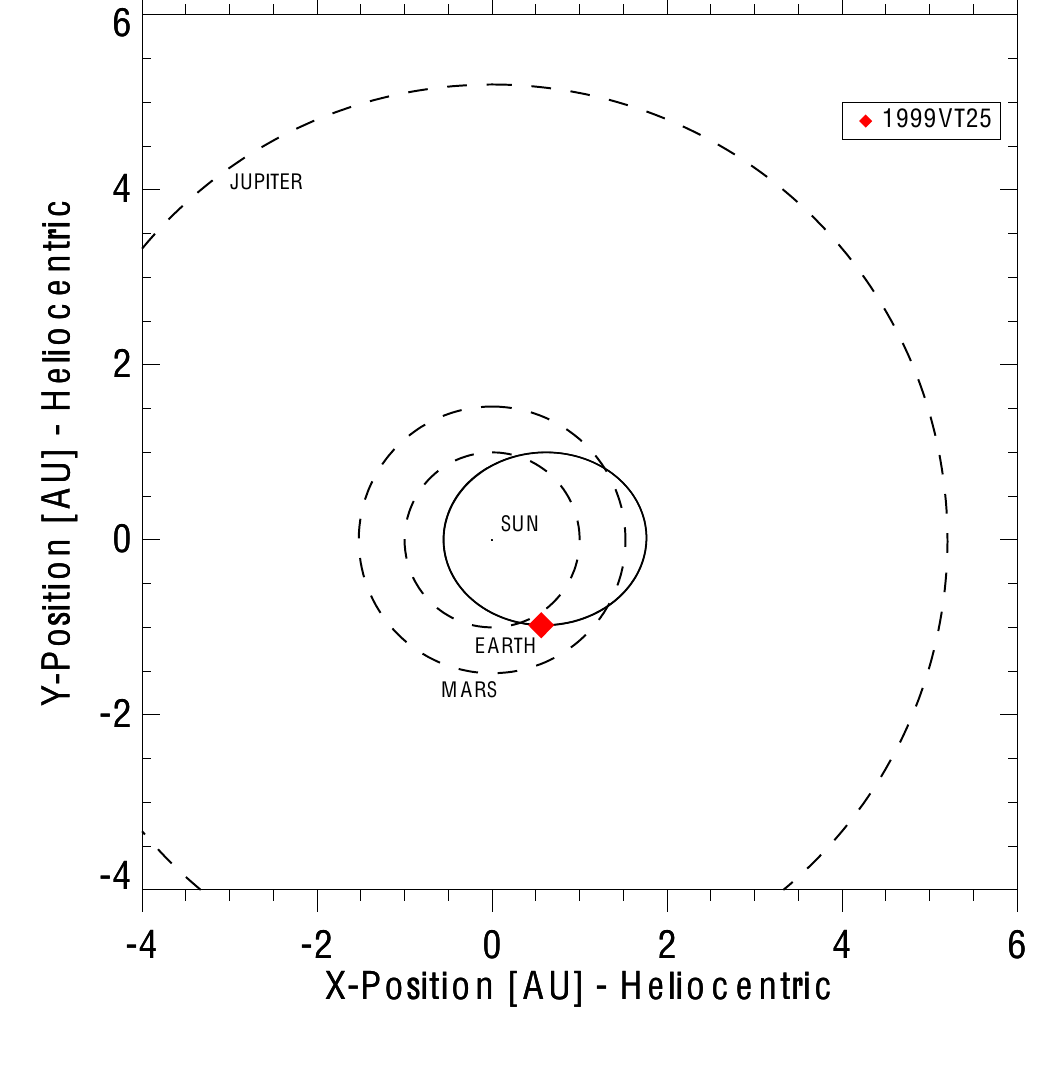}
   \includegraphics[width=0.65\columnwidth,trim=3.5cm 0.5cm 7.5cm 16.5cm,clip=true]{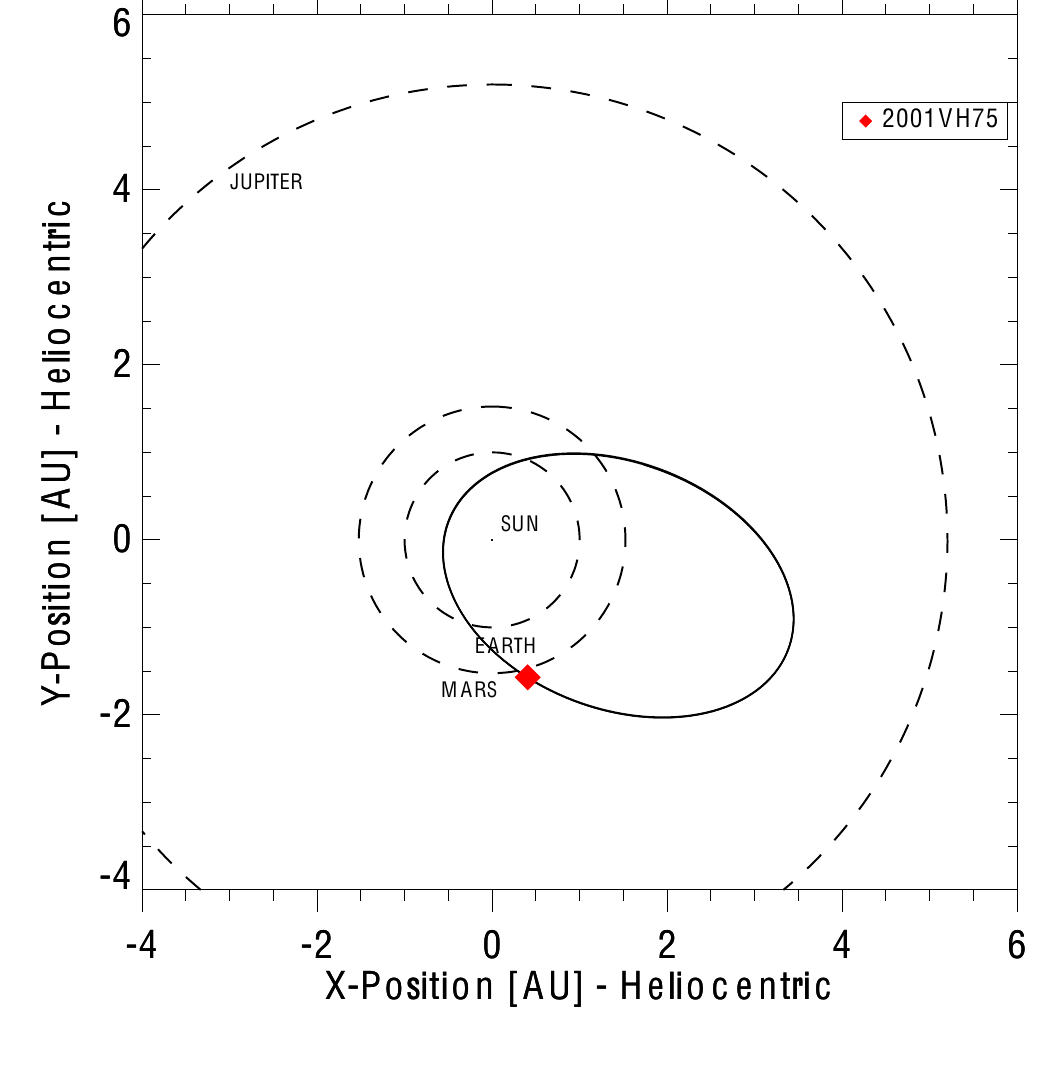}
   \includegraphics[width=0.65\columnwidth,trim=3.5cm 0.5cm 7.5cm 16.5cm,clip=true]{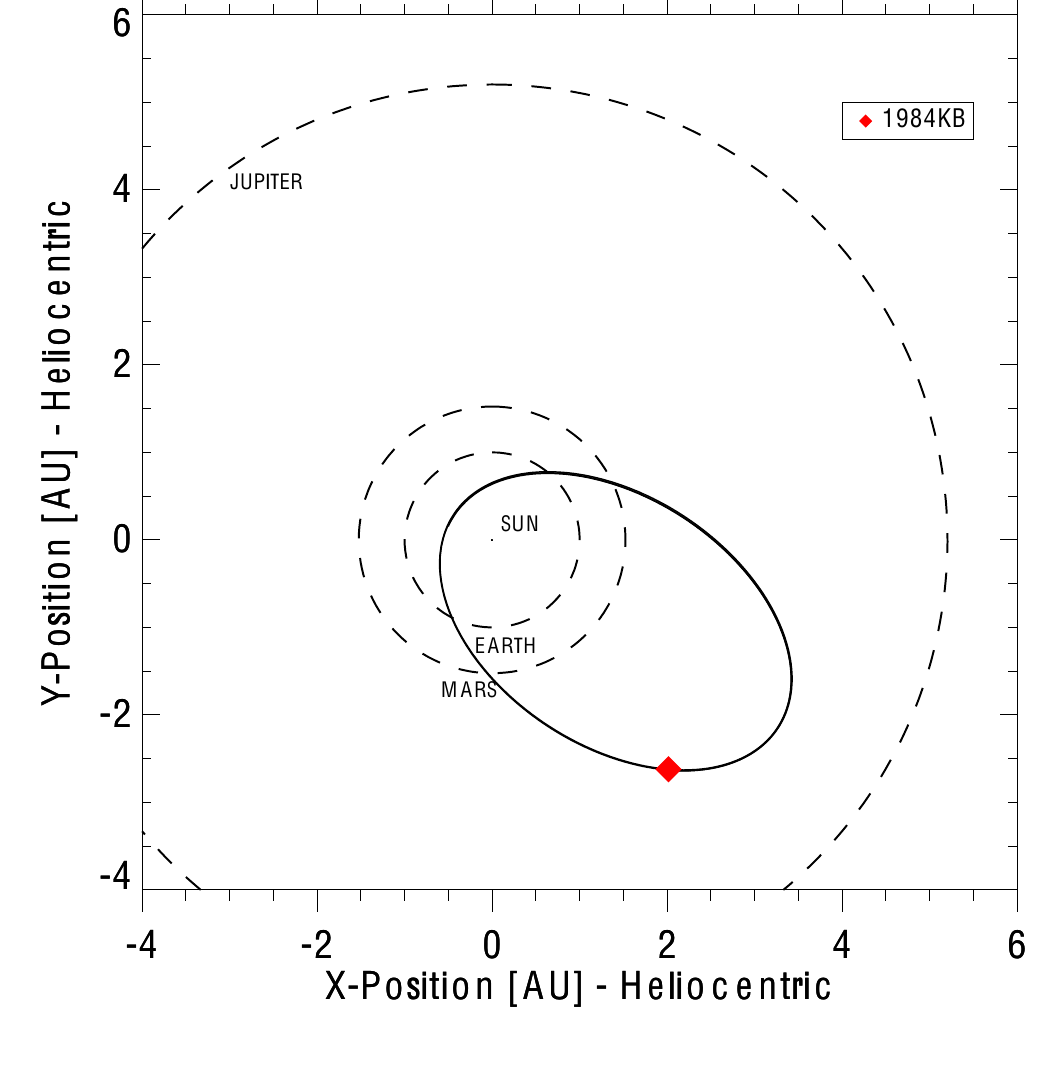}
   \includegraphics[width=0.65\columnwidth,trim=3.5cm 0.5cm 7.5cm 16.5cm,clip=true]{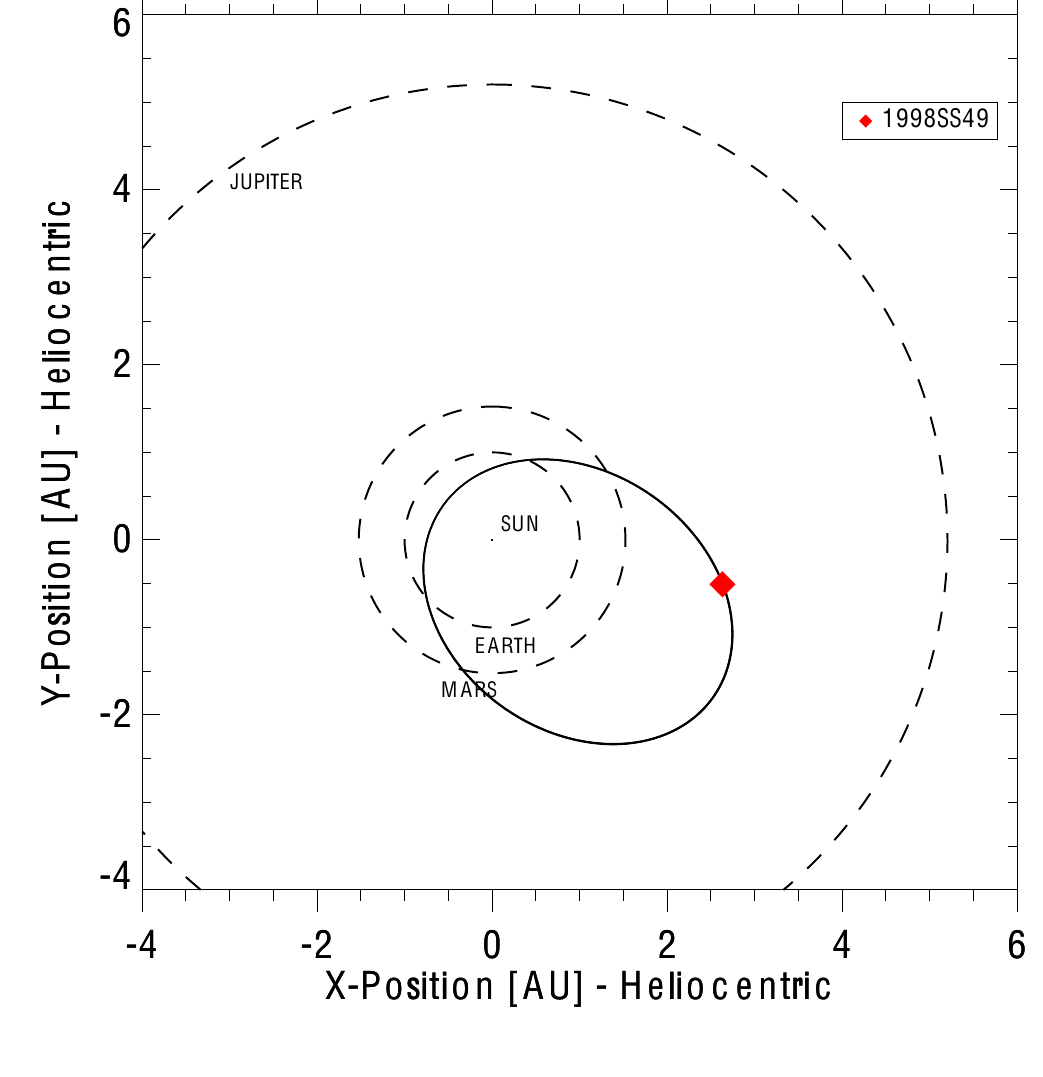}
   \includegraphics[width=0.65\columnwidth,trim=3.5cm 0.5cm 7.5cm 16.5cm,clip=true]{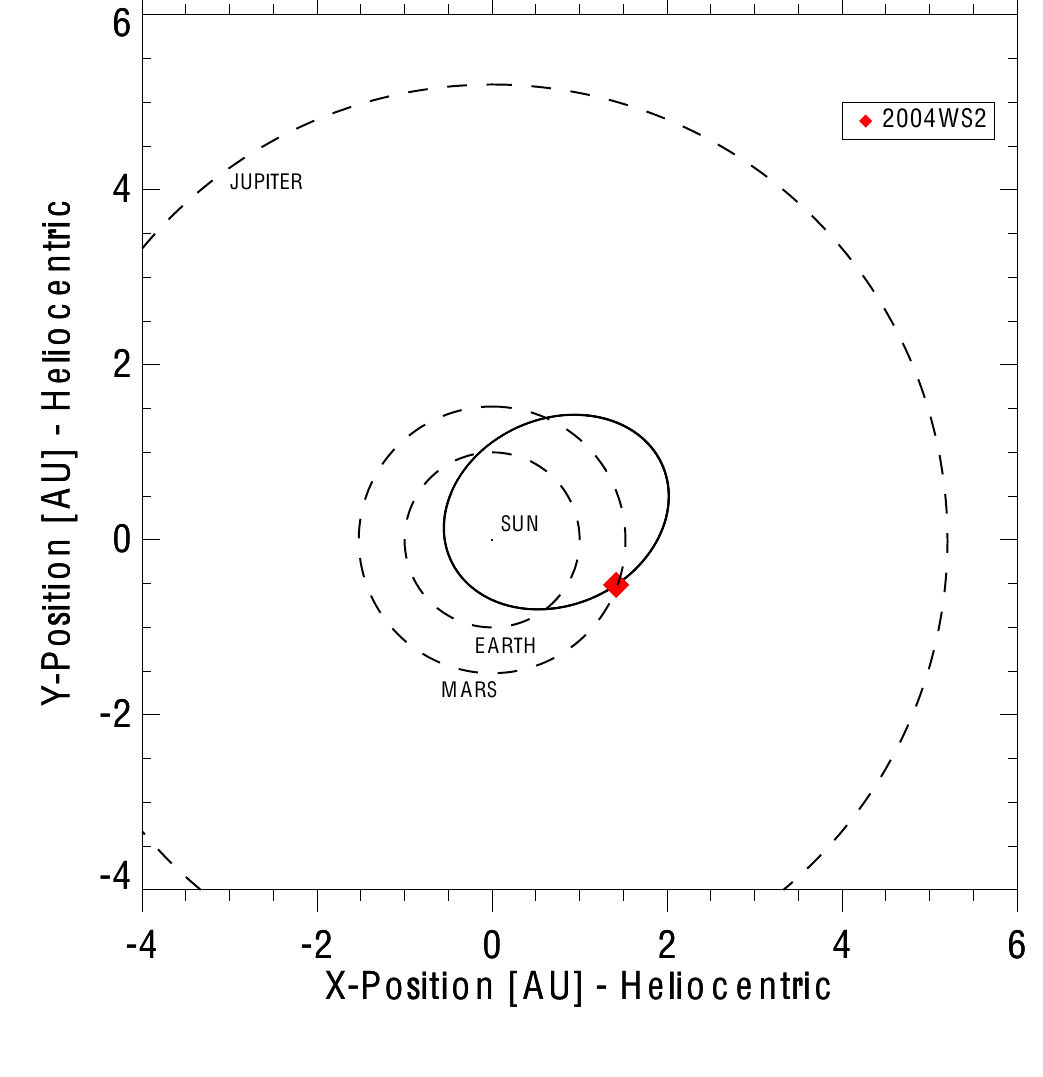}
   \includegraphics[width=0.65\columnwidth,trim=3.5cm 0.5cm 7.5cm 16.5cm,clip=true]{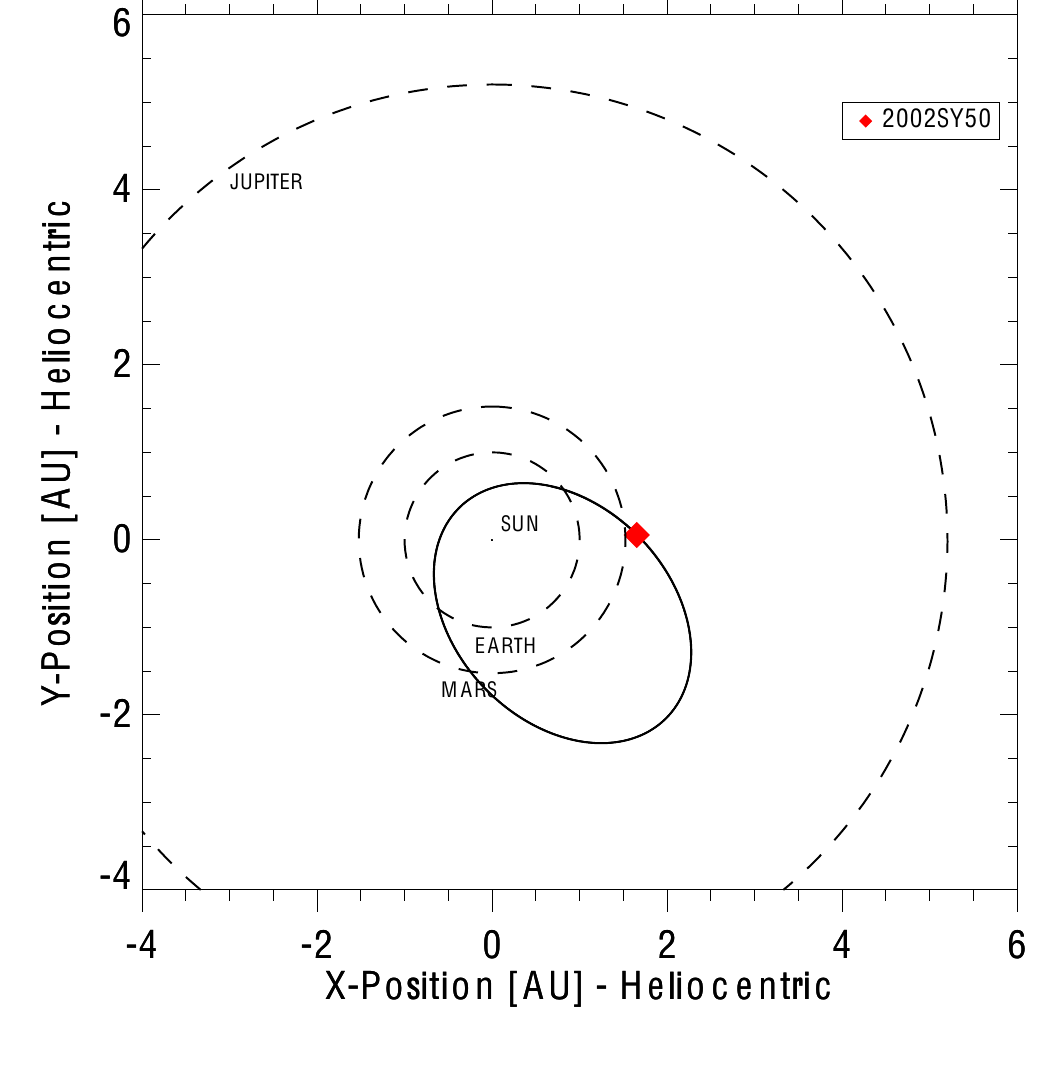}
   \caption{Orbital position of 2P/Encke and the NEOs at the time of the observations. The orbits of the Earth, Mars, and Jupiter are displayed.}
   \label{Orb-plot}
\end{figure*}

The NEOs to be observed were selected in the following way. 
\begin{enumerate}
\item A list was compiled of all the NEOs that have present day osculating orbits that have been shown by other authors to be similar to the mean orbit of the Taurids using one of the standard test listed in \citet{Jopek2013MNRAS}. 
\item This list was reduced by requiring that the NEOs were observable during the period of our observing run. With one exception, we also confined selection to those that have been numbered and so have secure orbits. 
\item From this reduced list, those NEOs that have been claimed by other authors to satisfy the criterion of \citet{Porubcan2004}, namely that the orbits have remained similar for 5000 years, were selected. These were 1998~QS$_{52}$, 1999~RK$_{45}$, 2003~QC$_{10}$, and 2003~UL$_{3}$. We refer to these NEOs as "Group 1" objects.
\item There was telescope time available to observe more NEOs than these 4 and so some were selected from those satisfying condition 2 above that have been officially numbered and so have secure orbits. These were 1984~KB, 1998~SS$_{49}$, 1999~VT$_{25}$, 2001~VH$_{75}$, 2002~SY$_{50}$, and 2004~WS$_{2}$. We refer to these NEOs as "Group 2" objects.
\end{enumerate}

In addition we observed two main belt asteroids (2000~CT$_{33}$ and 2004~RP$_{191}$) that are not related to 2P/Encke based on their orbits (see orbital elements in Tab. \ref{obs-table}). Numbers and names, where they exist, are given for each asteroid in Table \ref{targ-table}, but provisional designations are used throughout this paper.

\begin{table*}
\caption{Observational circumstances.}
\label{obs-table}
\begin{center}
\begin{tabular}{lccccccccc}
\hline
			&				&					&									& PHOTOMETRY	& SPECTROSCOPY \\

Object name	&	r$^{a}$ (AU)	& $\Delta$$^{b}$ (AU)	&	$\alpha$$^{c}$ (\degr)	&	t (s) x N$^{d}$ 	&	t (s) x N$^{e}$	& e$^{f}$ (\degr)	& i$^{g}$ (\degr)	& a$^{h}$ (AU)	& q$^{i}$ (AU)\\
\hline
2P/Encke		&	3.65		&2.71			&	6.98				&	180 x 3 &			360 x 3	& 0.85	& 11.78	&	2.21	& 0.34	\\
\hline
1998~QS$_{52}$	&	2.90		&2.19			&	16.44			&	180 x 5 &			180 x 4	& 0.86	& 17.56	&	2.20	& 0.31	\\
1999~RK$_{45}$	&	2.11		&1.20			&	16.57			&	180 x 4 + 60 x 3 &	1200 x 3	& 0.77	& 5.89	&	1.60	& 0.36	\\
2003~QC$_{10}$	&	1.52		&0.78			&	37.64			&	180 x 4 &			210 x 4	& 0.73	& 5.04	&	1.37	& 0.37	\\
2003~UL$_{3}$		&	3.10		&2.12			&	6.55				&	180 x 4 + 60 x 6 &	1200 x 3	& 0.80	& 14.66	&	2.25	& 0.46	\\
\hline
1999~VT$_{25}$	&	1.14		&0.23			&	52.85			&	11 x 5 + 5 x 5 	&	200 x 3	& 0.52	& 5.15	&	1.16	& 0.55	\\
2001~VH$_{75}$	&	1.63		&0.84			&	31.16			&	60 x 3 + 20 x 2 &	200 x 3	& 0.74	& 10.62	&	2.10	& 0.55	\\
1984~KB		&	3.32		&2.31			&	1.83				&	120 x 3 + 20 x 2 &	200 x 3	& 0.76	& 4.85	&	2.22	& 0.53	\\
1998~SS$_{49}$	&	2.71		&2.03			&	18.56			&	180 x 4 + 20 x 1&	380 x 3	& 0.64	& 10.76	&	2.37	& 0.69	\\
2004~WS$_{2}$		&	1.54		&0.85			&	38.14			&	180 x 3 &			200 x 3	& 0.60	& 8.26	&	1.34	& 0.53	\\
2002~SY$_{50}$	&	1.65		&1.30			&	37.93			&	180 x 4 + 30 x1 &	200 x 4	& 0.69	& 8.75	&	1.71	& 0.53	\\
\hline
2000~CT$_{33}$	&	2.78		&1.82			&	8.10				&	60 x 3 + 10 x 2 &	70 x 3	& 0.03	& 1.28	&	2.83	& 2.76	\\
2004~RP$_{191}$	&	2.11		&1.14			&	10.94			&	60 x 3 + 10 x 2 &	70 x 3	& 0.13	& 6.23	&	2.39	& 2.08	\\
\hline
\end{tabular}
\end{center}
$^{a}$Heliocentric distance (AU).
$^{b}$Geocentric distance (AU).
$^{c}$Phase angle (\degr).
$^{d}$Photometry: Exposure time (s) x Number of exposures.
$^{e}$Spectroscopy: Exposure time (s) x Number of exposures.
$^{f}$Eccentricity.
$^{g}$Inclination.
$^{h}$Semi-major axis.
$^{i}$Perihelion distance.
\end{table*}

\begin{table}
   \caption{Asteroids observed.}
   \label{targ-table}
   \begin{center}
   \begin{tabular}{l l l} 
\hline
Designation & Number & Name \\
\hline
1998~QS$_{52}$	& 16960	& -- \\
1999~RK$_{45}$	& 162195	& --	\\
2003~QC$_{10}$	& 405212	& --	\\
2003~UL$_{3}$		& 380455	& --	\\
\hline
1999~VT$_{25}$	& 152828	& --	\\
2001~VH$_{75}$	& 153792	& --	\\
1984~KB			& 6063	& Jason	\\
1998~SS$_{49}$	& 85713	& --	\\
2004~WS$_{2}$	& 170903	& --	\\
2002~SY$_{50}$	& 154276	& --	\\
\hline
2000~CT$_{33}$	& 16204	& --	\\
2004~RP$_{191}$	& 198008	& --	\\
\hline
   \end{tabular}
   \end{center}
\end{table}

\subsection{Photometry}
We observed 2P/Encke and the NEOs using the R Special filter ($\lambda_{central}$ = 638 nm). The exposure times and the number of exposures per object are summarised in Tab. \ref{obs-table}. Bias, flat fields and photometric standard star fields were taken during the night. The images were reduced using standard techniques of bias subtraction, flat field correction, exposure time normalisation and sky background subtraction. For the flux calibration we determined the zero point from the standard star fields observed during the night, assuming standard values for the extinction and colour coefficients, which do not add a significant uncertainty for a site like Paranal. The night was photometric. 

\subsection{Spectroscopy} Reflectance spectra of 2P/Encke and the NEOs were obtained using a 1.0\arcsec slit together with the low-resolution grism 150I+27, that covers the full visible spectrum from 400 nm to 900 nm on the detector. For calibration purposes, spectra of solar analogue stars were taken at similar airmasses and with the instrument in the same configuration. Standard methods of spectroscopy reduction and extraction were applied: bias subtraction, flat fielding using lamp flats, wavelength calibration using arc lamp spectra, exposure time normalisation, and sky background subtraction. To remove the solar spectrum from each of the objects, we divided the object's spectra by the average of the solar analogue spectra. The object spectra were then normalised to unity at 638 nm (central wavelength of the R special filter). 

\section{Results: 2P/Encke and the NEOs}
If 2P/Encke, the NEOs, Sutter's Mill and Maribo are parts of a same body we expect that they show, in addition to common dynamical properties, also similar spectral properties. Moreover, it is reasonable to expect that, since 2P/Encke shows activity in a good part of its orbit, the NEOs could show cometary activity as well. At the time of the observation, the majority of the NEOs were at an heliocentric distance where 2P/Encke has shown to be active. 
Figure \ref{Orb-plot} shows the orbits of 2P/Encke and the NEOs and their location at the time of the observations.

\subsection{Photometry}
\begin{figure}[h]
\centering
\includegraphics[width=0.95\columnwidth,trim=0cm 9.5cm 7cm 0cm,clip=true]{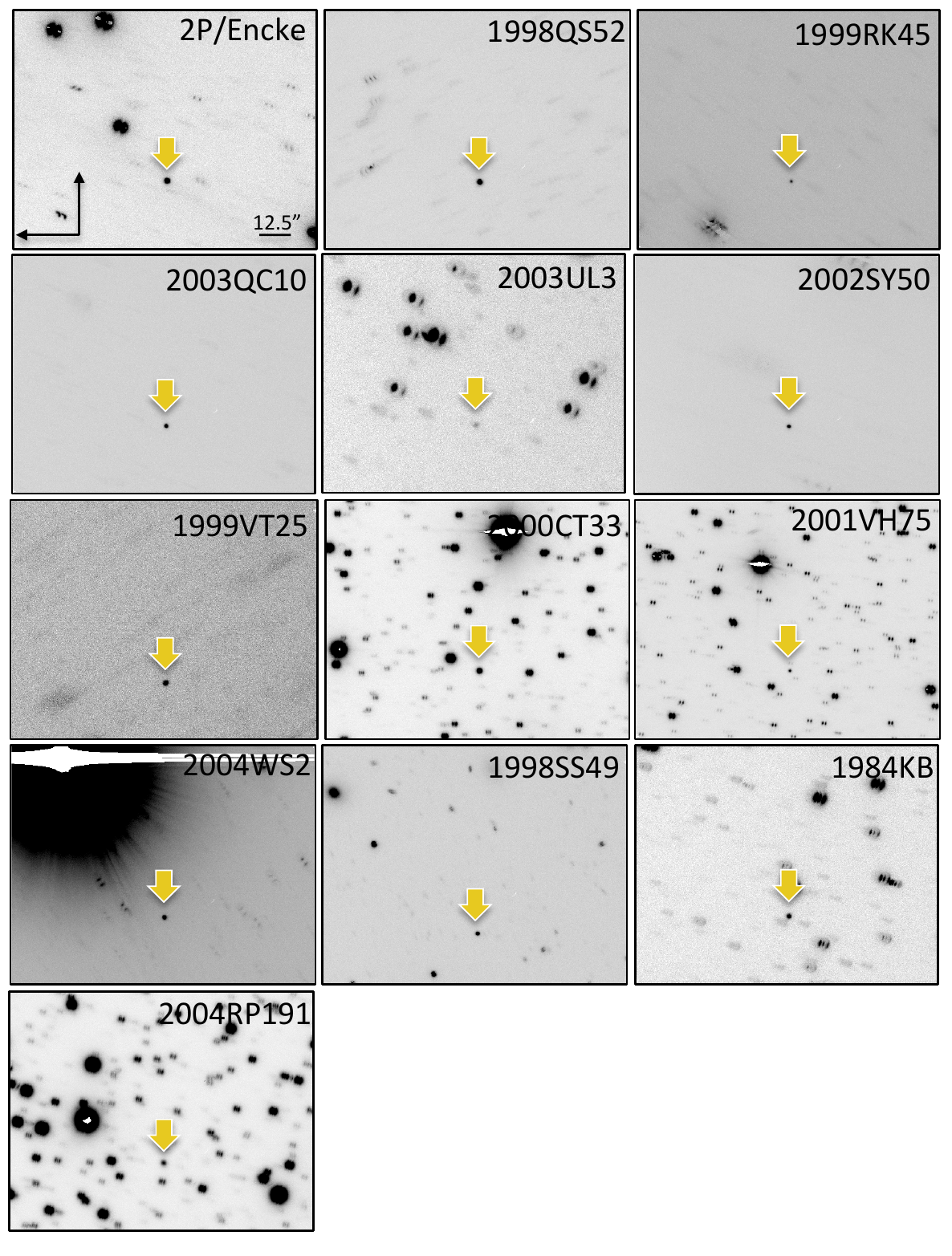}
\caption{Median combination of images of 2P/Encke and the NEOs. Image scale and orientation are shown. The position of the objects is marked by an arrow.}
   \label{Images-example}
\end{figure}

Examples of images of 2P/Encke and the NEOs are shown in Fig. \ref{Images-example}. 

Using aperture photometry we measured the magnitude of the object in each frame. The measured photometry was reduced to heliocentric distance r = 1 AU, geocentric distance $\Delta$ = 1 AU, and phase angle $\alpha$ = 0\degr  using:
\begin{equation}
H = m(1,1,0) = m_{meas} - 5log(r\Delta) + 2.5log[\Phi(\alpha)]
\end{equation}
\noindent where m$_{meas}$ is the measured magnitude. \newline
For the phase function correction we used a linear approximation (typically used for comets):
\begin{equation}
2.5log[\Phi(\alpha)] = - \alpha \cdot \beta
\end{equation}
\noindent with the linear phase coefficient $\beta$ = 0.05 mag/\degr as determined for 2P/Encke by \citet{Fernandez2000ic}, and the IAU-adopted phase law (typically used for asteroids) \citep{Meeus1998}:
\begin{equation}
2.5log[\Phi(\alpha)] = 2.5log[(1-G)\Phi_1 + G\Phi_2]
\label{eq_IAU_phasefunction}
\end{equation}
with
\begin{eqnarray}
\Phi_1 &=& exp\bigg[-3.33\bigg(tan\frac{\alpha}{2}\bigg)^{0.63}\bigg] \\
\Phi_2 &=& exp\bigg[-1.87\bigg(tan\frac{\alpha}{2}\bigg)^{1.22}\bigg]
\label{eqphi12}
\end{eqnarray}
with the ``slope parameter'' $G$ = 0.15 (a typical value for asteroids).

\begin{table}
\caption{Photometry results.}
\label{phot-results}
\begin{center}
\begin{tabular}{lccc}
\hline
Object name	&	H$_{a}$$^{a}$ (mag) 	&	H$_{a}$$^{b}$ (mag)	& R$_{eff}$ (km)	\\
\hline
2P/Encke		&	{\bf{14.17}} $\pm$	0.03		&	{\bf{14.34}} $\pm$	0.03		& 3.7	$^{c}$, 3.4$^{d}$	\\
\hline
1998~QS$_{52}$	&	14.03 $\pm$	0.03		&	14.09 $\pm$	0.03		& 3.9$^{c}$, 3.8$^{d}$	\\
1999~RK$_{45}$	&	18.77 $\pm$	0.05		&	18.83 $\pm$	0.05		& 0.4$^{c}$, 0.4$^{d}$\\
2003~QC$_{10}$	&	17.61 $\pm$	0.03		&	17.24 $\pm$	0.03		& 0.8$^{c}$, 0.9$^{d}$\\
2003~UL$_{3}$		&	17.46 $\pm$	0.08		&	17.63 $\pm$	0.08		& 0.8$^{c}$, 0.7$^{d}$\\
\hline
1999~VT$_{25}$	&	20.98 $\pm$	0.05		&	20.28 $\pm$	0.05		& 0.2$^{c}$, 0.2$^{d}$\\
2001~VH$_{75}$	&	18.08 $\pm$	0.04		&	17.86 $\pm$	0.04		& 0.6$^{c}$, 0.7$^{d}$\\
1984~KB		&	15.87 $\pm$	0.04		&	16.00 $\pm$	0.04		& 1.7$^{c}$, 1.6$^{d}$	\\
1998~SS$_{49}$	&	15.46 $\pm$	0.04		&	15.49 $\pm$	0.04		& 2.0	$^{c}$, 2.0$^{d}$\\
2004~WS$_{2}$		&	17.74 $\pm$	0.03		&	17.36 $\pm$	0.03		& 0.7$^{c}$, 0.8$^{d}$\\
2002~SY$_{50}$	&	17.01 $\pm$	0.04		&	16.64 $\pm$	0.04		& 1.0$^{c}$, 1.2$^{d}$\\
\hline
2000~CT$_{33}$	&	14.26 $\pm$	0.03		&	14.42 $\pm$	0.03		& 3.5$^{c}$, 3.3$^{d}$\\
2004~RP$_{191}$	&	16.45 $\pm$	0.04		&	16.58 $\pm$	0.04		& 1.3$^{c}$, 1.2$^{d}$\\
\hline
\end{tabular}
\end{center}
$^{a}$Average absolute magnitude of the object and associated error obtained using the IAU-adopted phase law as phase function approximation.
$^{b}$Average absolute magnitude of the object and associated error obtained using the linear phase function approximation.
$^{c}$ Effective radius corresponding to the absolute magnitude in column a.
$^{d}$ Effective radius corresponding to the absolute magnitude in column b. 
\end{table}

The average absolute magnitude obtained with both phase function approximations is summarised in Tab. \ref{phot-results}. The corresponding effective radius of the objects (in km) is calculated using \citep{Meeus1998}:
\begin{equation}
R_{eff} = 1.496 \times 10^8 \sqrt{\frac{10^{0.4(M_{\sun}-H})}{A}}
\end{equation}
where $A=0.04$ is the assumed geometric albedo of the object, $H$ and $M_{\sun}$ are the absolute magnitude of the object and of the Sun in the same wavelength, respectively.\newline

If the Taurid NEOs are indeed pieces of a parent body that produced 2P/Encke, it is possible that their albedo is low. Primitive and low-albedo objects may have values of the "slope parameter" G much lower than the one generally assumed for asteroids. Table \ref{slope-param} summarises the average absolute magnitudes of the observed objects for G=0.05 and G=-0.25. The first value is representative of low-albedo objects and the second one is the slope parameter determined by \citet{Fernandez2000ic} for 2P/Encke.

\begin{table}
\caption{Average absolute magnitude of the observed objects determined with different values of the "slope parameter".}
\label{slope-param}
\begin{center}
\begin{tabular}{lccc}
\hline
Object name	&H$_{G=0.15}$ (mag) & H$_{G=0.05}$ (mag) & H$_{G=-0.25}$ (mag)\\
\hline
2P/Encke			& 	14.17$\pm$0.03 	& 14.10	& 13.87 \\
\hline
1998~QS$_{52}$	&	14.03$\pm$ 0.03	& 13.91 	& 13.44 \\
1999~RK$_{45}$	&	18.77$\pm$0.05	& 18.65	& 18.18 \\
2003~QC$_{10}$	&	17.61$\pm$0.03	& 17.41	& 16.37 \\
2003~UL$_{3}$	&	17.46$\pm$0.08	& 17.39 	& 17.17 \\
\hline
1999~VT$_{25}$	&	20.98$\pm$0.05	& 20.74	& 19.25 \\
2001~VH$_{75}$	&	18.08$\pm$ 0.04	& 17.90	& 17.05 \\
1984~KB			&	15.87$\pm$0.04	& 15.84	& 15.75 \\
1998~SS$_{49}$	&	15.46 $\pm$0.04	& 15.33	& 14.81 \\
2004~WS$_{2}$	&	17.74$\pm$0.03	& 17.54	& 16.49 \\
2002~SY$_{50}$	&	17.01$\pm$0.004 	& 16.81	& 15.77 \\
\hline
\end{tabular}
\end{center}
{\it{Note}}: The error associated to the average absolute magnitude, as reported in the second column, is representative of the uncertainty associated to all determinations of the average absolute magnitude.
\end{table}

\begin{figure*}
\centering
\includegraphics[width=1.0\textwidth,trim=3cm 0.5cm 0.6cm 0cm,clip=true]{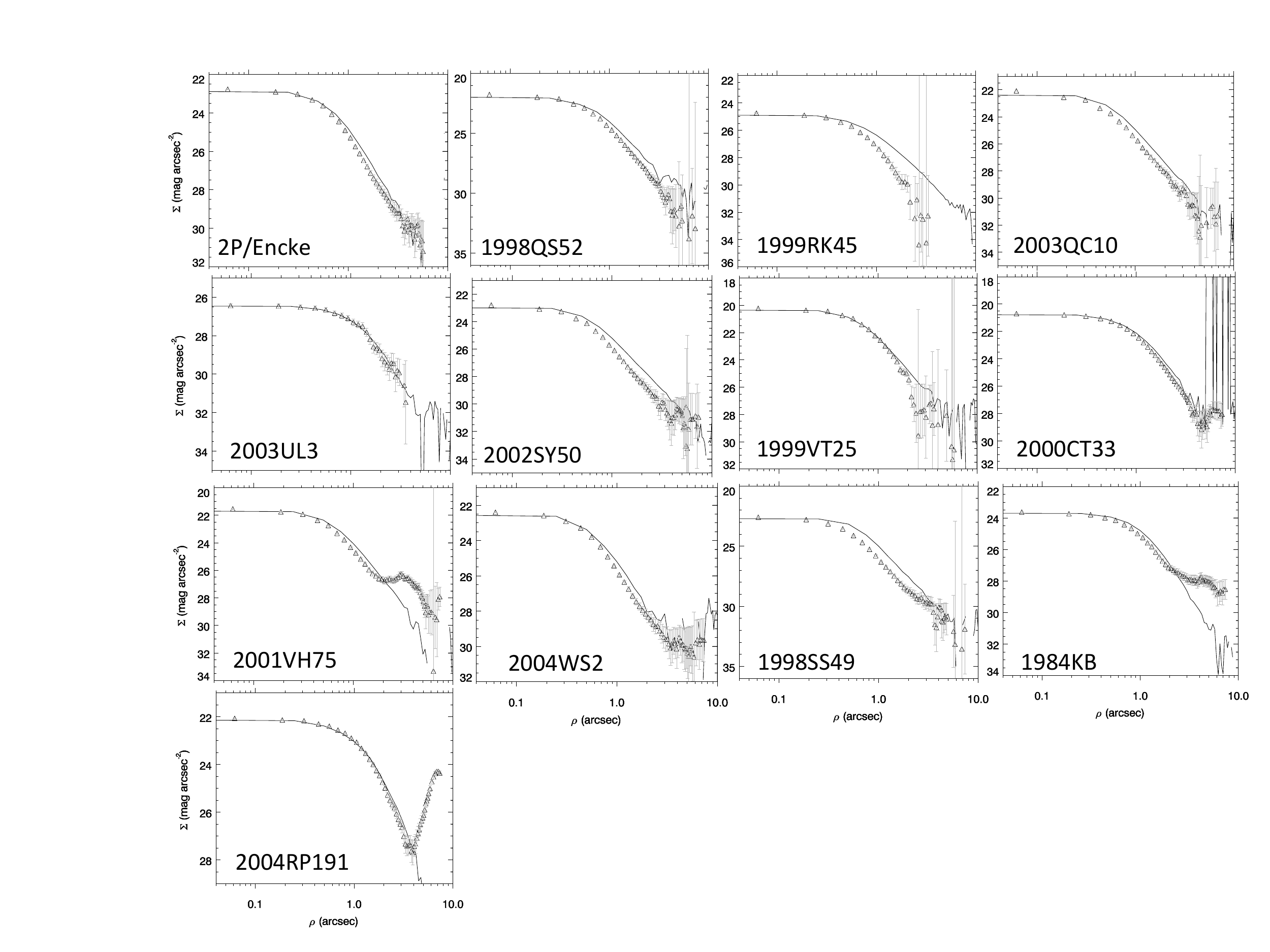}
\caption{Surface brightness profiles for 2P/Encke and the NEOs. For each object, the triangles show the profile of the object, while the solid line shows the image PSF. }
   \label{Radial-profiles}
\end{figure*}

To investigate the possible presence of cometary activity around the nucleus of 2P/Encke and the NEOs, we compared the surface brightness profile of the objects with the one of a star, following the method described in \citet{Tubiana2008aa}: we measure the profile across the width of the trailed star profile in the same image as the asteroid. All objects (including 2P/Encke) show star-like profiles (see Fig. \ref{Radial-profiles}), indicating that no detectable activity is present around the object.

However, 2P/Encke is brighter than expected. In fact, the effective radius that we obtained is larger then the one determined by \citet{Fernandez2000ic}. 
Here are a few possible explanations of this discrepancy:
\begin{enumerate}
\item 2P/Encke is weakly active at the time of our observations. No resolved coma is present around the nucleus, as shown by the radial profile (see Fig. \ref{Radial-profiles}), but it is possible that activity is within the seeing disc and therefore at a low level. Comparing the magnitude we observe with that found by  \citet{Fernandez2000ic}, we would expect that the coma contributes 58\% of the flux within the aperture.  \newline
However, it might not be reasonable to believe that about $\sim$ 60\% of the additional light is within a coma area of $\sim$ 3000 km around the nucleus and it drops in a way that it is not detected in the more distant profile, since the dust coma intensity drops following 1/r, while the seeing disc is steeper. Significant activity cannot be the only source of the discrepancy. 
\item  \citet{Fernandez2000ic} determined the radius of 2P/Encke using the flux corresponding to the midway between the maximum and the minimum flux of the rotational light curve. If our observation lays close to the maximum of the rotational light curve, our flux is higher than the one measured by \citet{Fernandez2000ic}, and this could contribute to the discrepancy in the radius.
\item The set of parameters used in the thermal model used to determine the radius of 2P/Encke contains various assumptions, such as the beaming parameter, which are reasonable but contribute some extra uncertainty.
\item When \citet{Fernandez2000ic} observed 2P/Encke, the comet was active. They estimated the coma contribution to the flux and they subtracted it. This determination might be source of the discrepancy.
\item The photometry of 2P/Encke obtained in this analysis is well in agreement with what was found by \citet{Fernandez2005} for similar heliocentric distances and phase angle. This could mean that the nucleus of 2P/Encke is indeed larger than the determination of \citet{Fernandez2000ic}, that the phase darkening is steeper and that the radar results obtained by \citet{HarmonNolan2005} need to be re-interpreted. Another explanation could be that 2P/Encke has a consistent activity in the orbital arc 3.5-3.7 AU outbound, although activity cannot completely explain the discrepancy, as discussed in point 1 above.
\end{enumerate}
The radial profiles of the NEOs rule out strong activity of these objects at the time of the observation, but they could still have an "Encke"-like behaviour, as described in scenario 1 above. Moreover, we have only snap-shot observations and each object was at a different location in its orbit, at different heliocentric distance. 2P/Encke, 2003~UL$_{3}$, 1999~VT$_{25}$, 2001~VH$_{75}$, 1984~KB, and 2004 WS2 were moving outbound while the other NEOs (1998~QS$_{52}$, 1999~RK$_{45}$, 2003~QC$_{10}$, 1998~SS$_{49}$, and 2002~SY$_{50}$) were in the inbound leg. Comets tend to show activity to larger distance post perihelion \citep{Kelley2013}.
Heliocentric light curves of the NEOs would provide a more complete information about the activity behaviour of these objects at different distances from the Sun. \newline
Assuming that the observed NEOs are truly inactive, a couple of objects are of comparable size as 2P/Encke. If they are all fragments of the same body, Encke's apparently unique continuing activity can not be explained by it being the largest fragment.

\subsection{Spectroscopy}
\label{sec-spectroscopy}
\begin{figure}
\centering
\includegraphics[width=0.95\columnwidth,trim=0.5cm 12.5cm 6.5cm 4cm,clip=true]{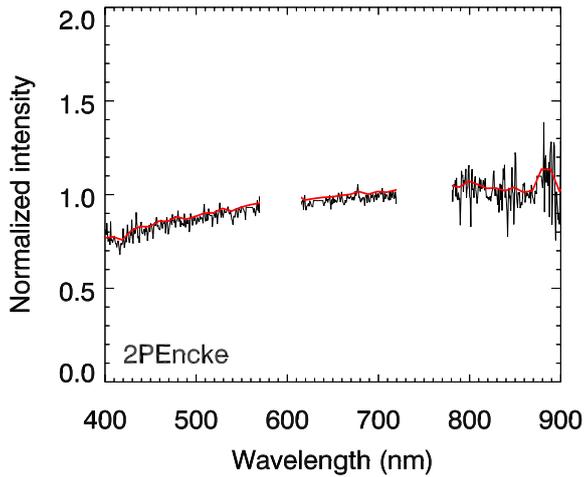}
\caption{Low resolution spectrum of the nucleus of 2P/Encke. The spectrum is normalised to unity at 638 nm. On the original spectrum (black line), the binned one (red line) is superimposed. The gaps around 600 nm and 750 nm are areas that have been masked due to artefacts created by issues in the solar analogue star spectrum at those wavelengths. The same normalisation, binning and masking also applies to all other spectra shown in this paper (Figures \ref{Spectra-NEOs1}, \ref{Spectra-NEOs2}, \& \ref{Spectra-NEOs3}).}
   \label{Spectrum-Encke}
\end{figure}

We obtained 3-4 low resolution spectra of each object in the wavelength range 400-900 nm. For each object, since all spectra looked very similar, we median them to increase the signal-to-noise ratio. An issue with the solar analogue stars creates an artefact in the spectra around 600 nm and around 750 nm, thus we masked these regions to avoid wrong interpretations of the spectra.\newline
Figure \ref{Spectrum-Encke} displays the low resolution spectrum of the nucleus of 2P/Encke. This is the first high signal-to-noise ratio visible spectrum of the bare nucleus of this comet. It is featureless and displays a reddening slope of (7.3 $\pm$ 0.2)\%/100 nm in the wavelength range $\Delta\lambda$ = 420 - 750 nm and (5.0 $\pm$ 0.2)\%/100 nm in the wavelength range $\Delta\lambda$ = 420 - 900 nm. The continuum of 2P/Encke flattens off beyond about 800 nm, which results in the shallower slope for the wider wavelength coverage.
As FORS spectra are noisy beyond 800 nm, we use the former range to measure the slope for all other objects.

Very few spectra of cometary nuclei exist in the literature, they are in general featureless and with a fairly constant and moderate slope in the visible wavelength range \citep{Luu1993ic,Tubiana2008aa,Tubiana2011}. Comet 67P/Churyumov-Gerasimenko has a featureless spectrum with a reddening slope of 11 $\pm$ 2 \%/100 nm in the same wavelength range, from FORS observations with an identical setup \citep{Tubiana2011}. Compared to the available spectra of other cometary nuclei, the spectrum of 2P/Encke is rather typical, despite 2P/Encke being in a very peculiar orbit.

\begin{table}[h]
\caption{Reddening slopes of low resolution spectra.}
\label{slope-spectra}
\begin{center}
\begin{tabular}{lc}
\hline
Object		& S$^a$ (\%/100nm)	\\
\hline
2P/Encke		& 7.3 $\pm$ 0.2		\\
			& 5.0 $\pm$ 0.2$^b$	\\
\hline
1998~QS$_{52}$	& 8.1$\pm$ 0.3		\\	
1999~RK$_{45}$	& 6.1 $\pm$ 0.6		\\
2003~QC$_{10}$	& 2.7 $\pm$ 0.4		\\
2003~UL$_{3}$		& 7 $\pm$ 2		\\
\hline
1999~VT$_{25}$	& -0.4 $\pm$ 0.4		\\
2001~VH$_{75}$	& 9.2 $\pm$ 0.4		\\
1984~KB		& 9.8 $\pm$ 0.5		\\
1998~SS$_{49}$	& 9.1 $\pm$ 0.3		\\
2004~WS$_{2}$		& 7.7 $\pm$ 0.3	\\
2002~SY$_{50}$	& 8.1 $\pm$ 0.3	\\
\hline
2000~CT$_{33}$	& 12.9 $\pm$ 0.3		\\
2004~RP$_{191}$	& 13.2 $\pm$ 0.4		\\
\hline
\end{tabular}
\end{center}
$^a$ Reddening slope over $\lambda$ = 420 -- 750 nm. \\
$^b$ Measured over $\lambda$ = 420 -- 900 nm.\\
\end{table}

\begin{figure}
\centering
\includegraphics[width=0.47\columnwidth,trim=0.1cm 15.0cm 9.5cm 5cm,clip=true]{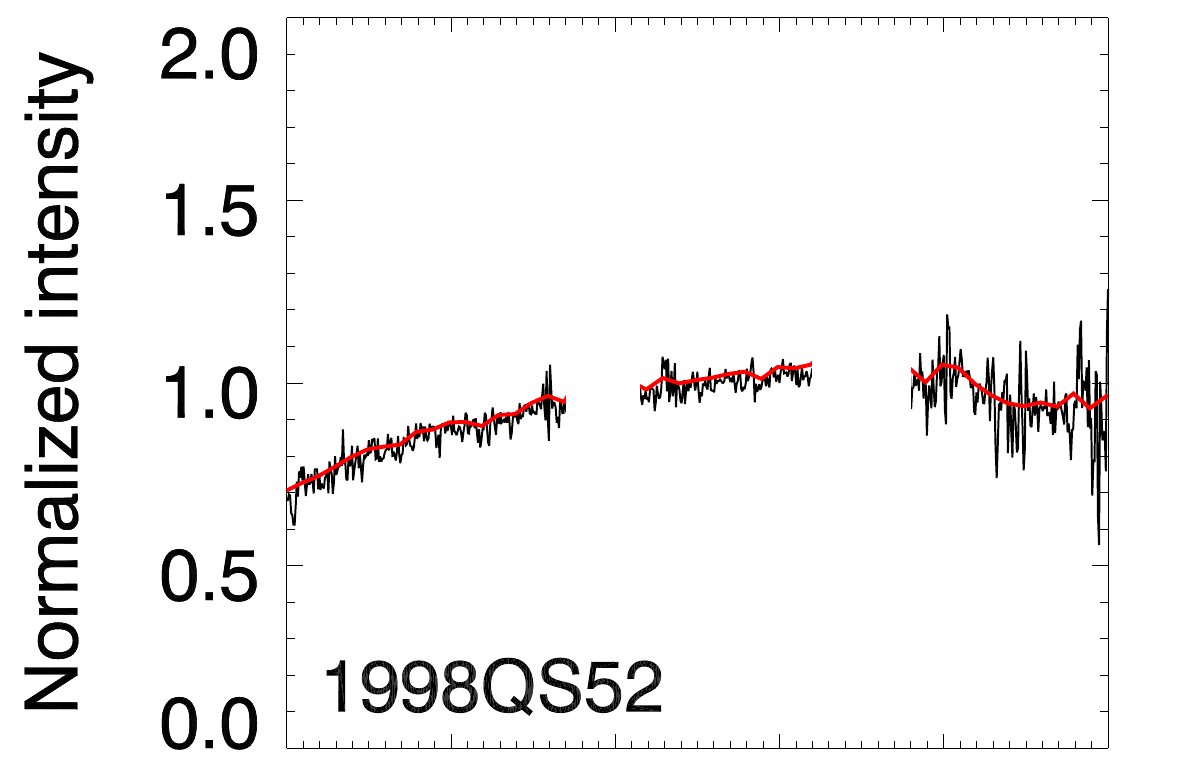}
\includegraphics[width=0.415\columnwidth,trim=1.5cm 15.0cm 9.6cm 5cm,clip=true]{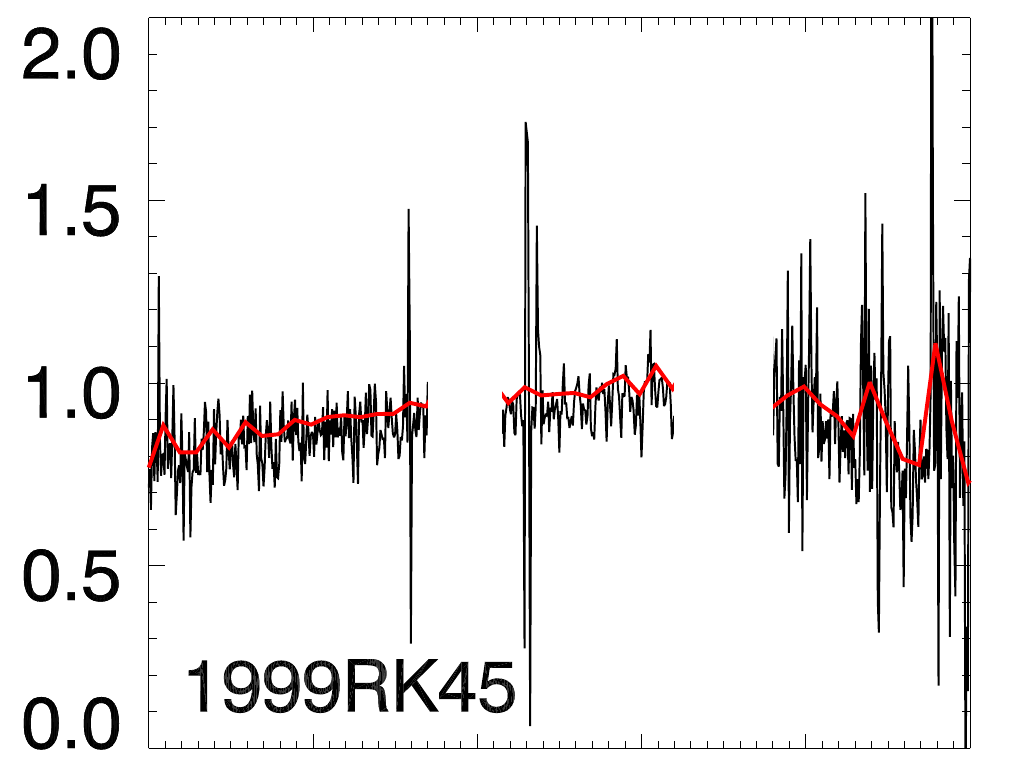}
\includegraphics[width=0.47\columnwidth,trim=0.1cm 12.5cm 9.5cm 5cm,clip=true]{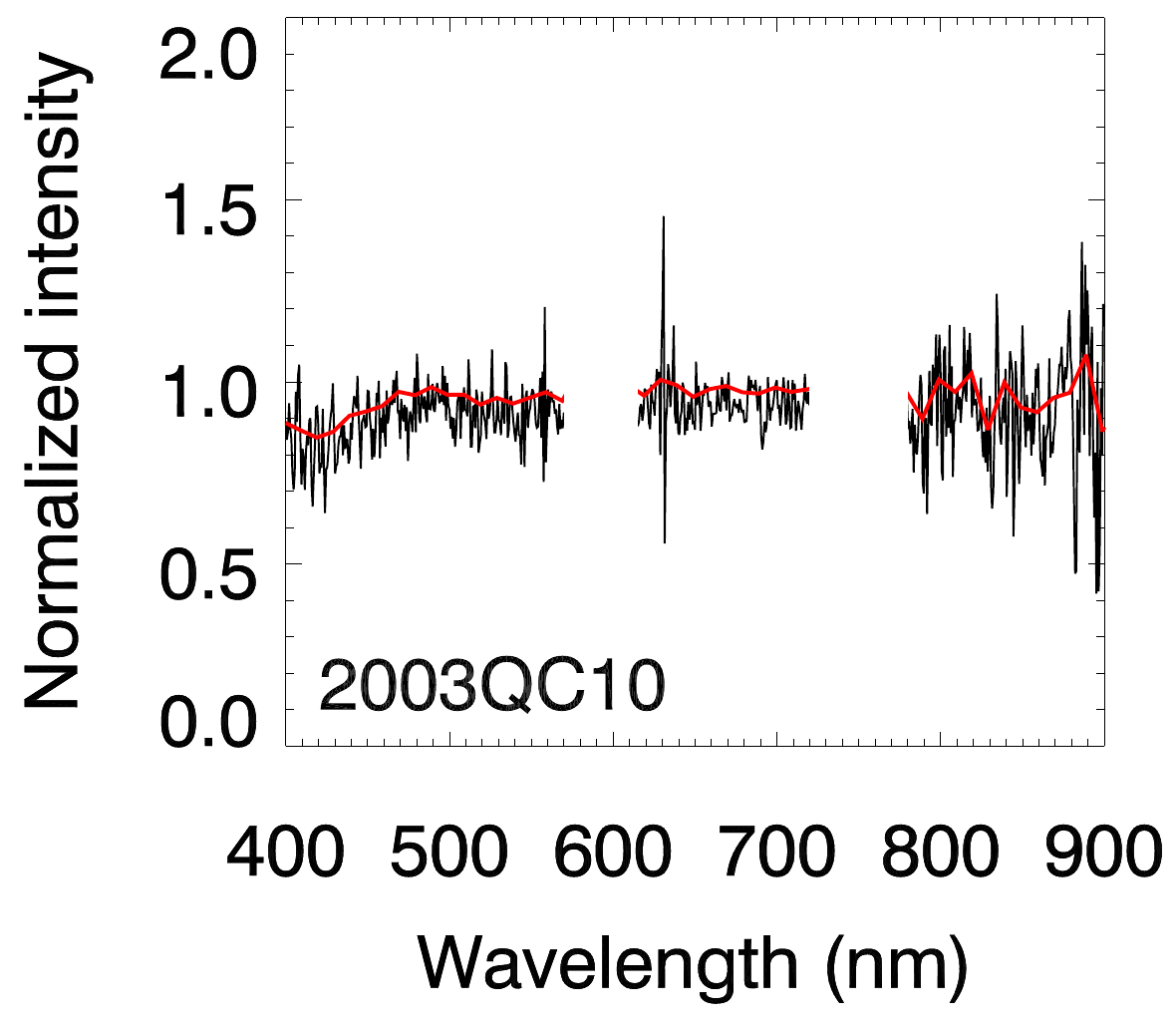}
\includegraphics[width=0.415\columnwidth,trim=1.5cm 12.5cm 9.6cm 5cm,clip=true]{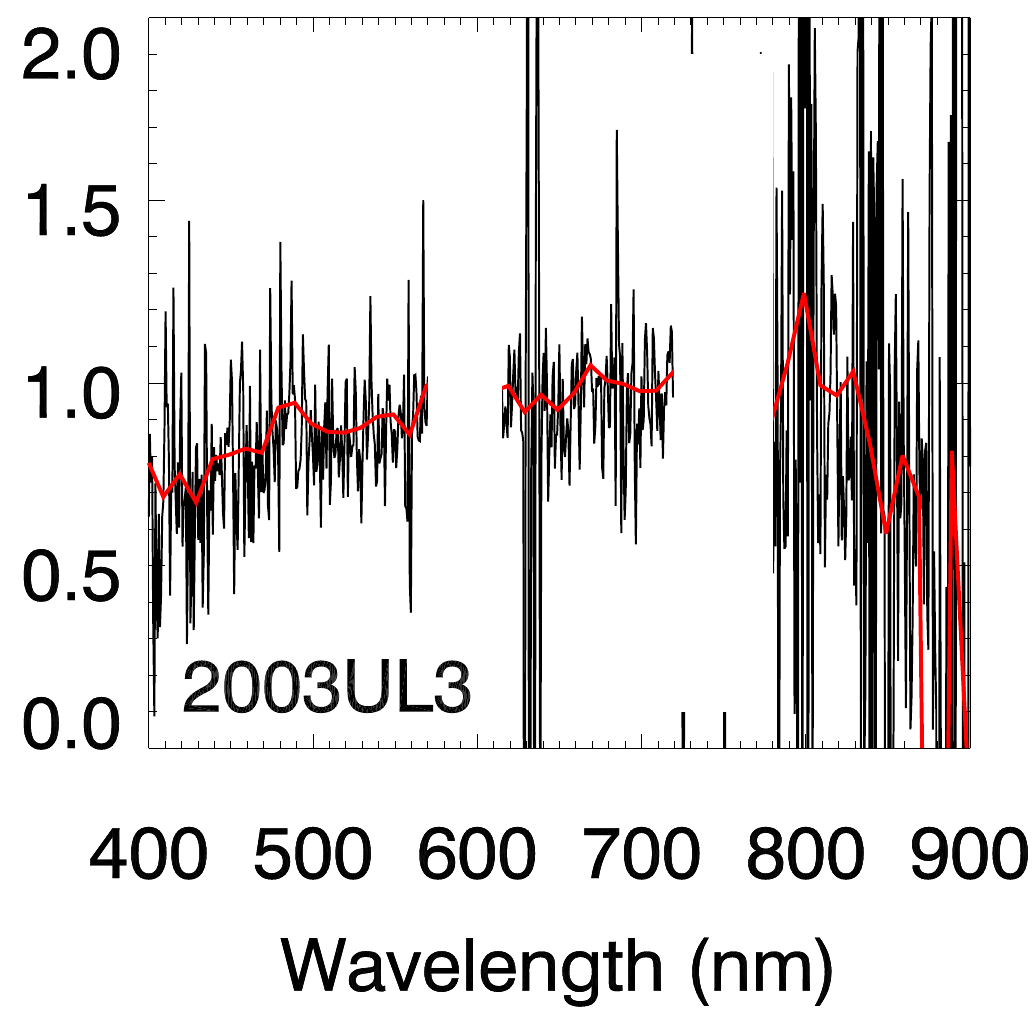}
\caption{Low resolution spectra of 1998~QS$_{52}$, 1999~RK$_{45}$, 2003~QC$_{10}$, and 2003~UL$_{3}$. The spike visible at about 650 nm is a skyline not completely removed during the spectra reduction.
}
\label{Spectra-NEOs1}
\end{figure}

The spectra of 1998~QS$_{52}$, 1999~RK$_{45}$, 2003~QC$_{10}$, and 2003~UL$_{3}$ (NEOs belonging to Group 1) are shown in Fig. \ref{Spectra-NEOs1}. All spectra look featureless for $\lambda$ < 800 nm and their reddening slopes are summarised in Table \ref{slope-spectra}. \citet{Michelsen2006} observed a strong phase reddening in NEO spectra at $\alpha$ > 60 \degr. Our observations were performed at $\alpha$ $\leq$ 53 \degr, thus the phase reddening should not hinder our comparison between asteroids.

\begin{figure}[h]
\centering
\includegraphics[width=0.95\columnwidth, trim=0.2cm 12.5cm 6.5cm 4cm,clip=true]{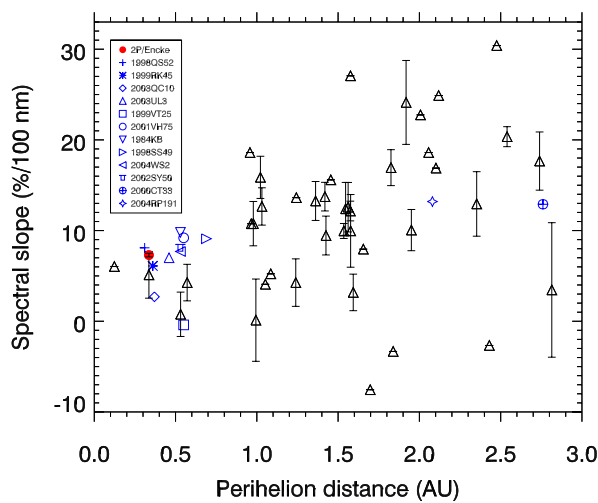}
\caption{Spectral slope versus perihelion distance, where a weak trend of steeper slopes at larger distances is found. Data taken from the MBOSS colour database \citep{Hainaut2012}. The red filled circle and the blue symbols correspond to the spectral slope of the spectrum of 2P/Encke and the observed asteroids, respectively, determined in this work (wavelength range $\Delta \lambda$ = 420 -- 750 nm).}
\label{spectralslope-vs-q}
\end{figure}

1998~QS$_{52}$, 1999~RK$_{45}$, and 2003~UL$_{3}$ display a slope comparable to the one of 2P/Encke. The spectrum of 2003~UL$_{3}$ is quite noisy, we do not consider the drop of flux at the red-end of the spectrum as a real feature, due to the low signal-to-noise ratio in this region. The spectrum of 1998~QS$_{52}$ has higher signal-to-nose ratio also at long wavelengths and the drop of flux at $\lambda >$ 800 nm seems real.  
2003~QC$_{10}$, instead, has a flatter spectrum than the nucleus of 2P/Encke. The integration back in time of its orbit supports the hypothesis that this NEO is a member of the family, but the orbit must have evolved. Based on the assumption on common origins, the possible explanations of the different slopes are:
\begin{enumerate}
\item A close approach to a planet might have produced changes in the surface properties of 2003~QC$_{10}$ and therefore its spectral slope \citep{DeMeo2014}.
\item Heating might have changed the surface of 2003~QC$_{10}$. 2P/Encke spends more time close to the Sun than 67P/Churyumov-Gerasimenko. Comparing the spectrum of 2P/Encke with the one of 67P/Churyumov-Gerasimenko, the first is slightly shallower ($\sim$ 7\%/100nm for 2P/Encke and $\sim$ 11\%/100nm for 67P/Churyumov-Gerasimenko). The orbit of 2003~QC$_{10}$ experiences greater average insolation over its orbit than 2P/Encke, thus higher heating might be the explanation of a shallower spectral slope.
\citet{Hainaut2012} found that minor bodies in the outer solar system show a weak trend of steeper slopes at larger distances, as shown in Fig. \ref{spectralslope-vs-q}. 
\item Another possibility is that the giant comet collided with another object, causing an initial fragmentation so that 2003~QC$_{10}$ is essentially from another body (the remains of the impactor). This is a possible explanation of why 2003~QC$_{10}$ is different from Encke and the other 3 asteroids belonging to Group 1, although it seems improbable.
\end{enumerate}

\begin{figure}[h]
\centering
\includegraphics[width=0.47\columnwidth,trim=0.1cm 15.0cm 9.5cm 5cm,clip=true]{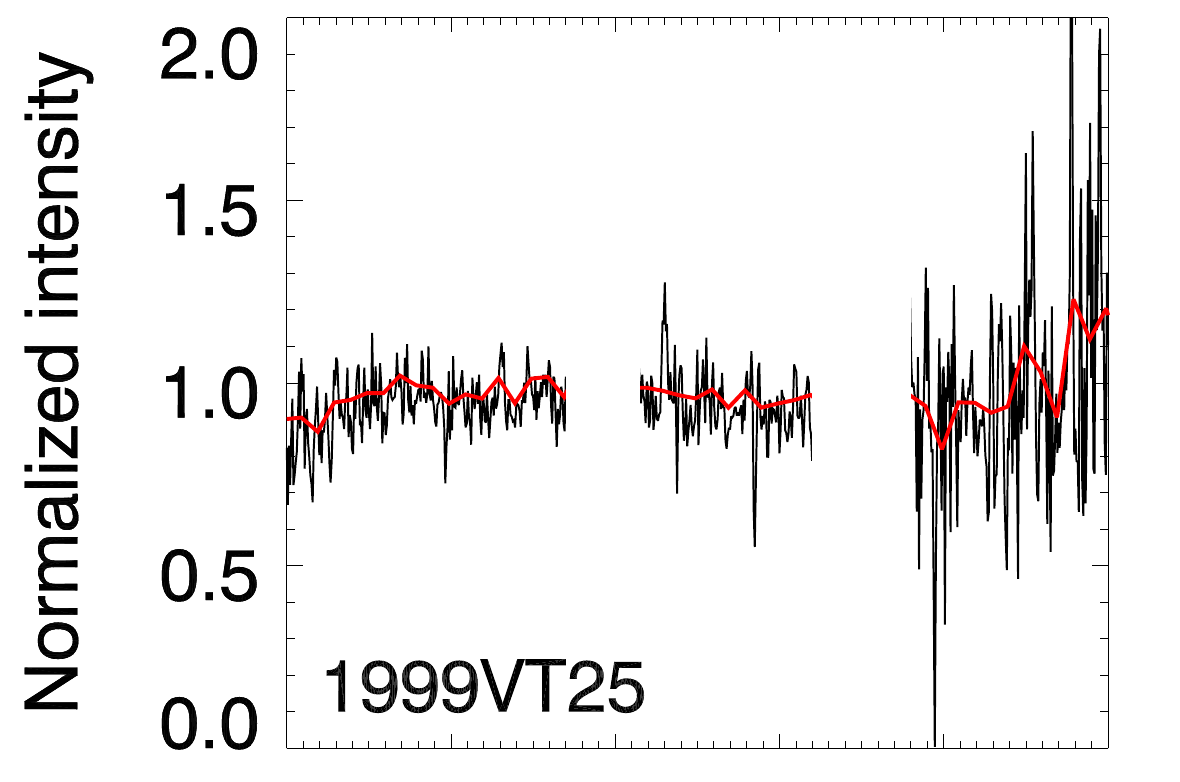}
\includegraphics[width=0.415\columnwidth,trim=1.5cm 15.0cm 9.6cm 5cm,clip=true]{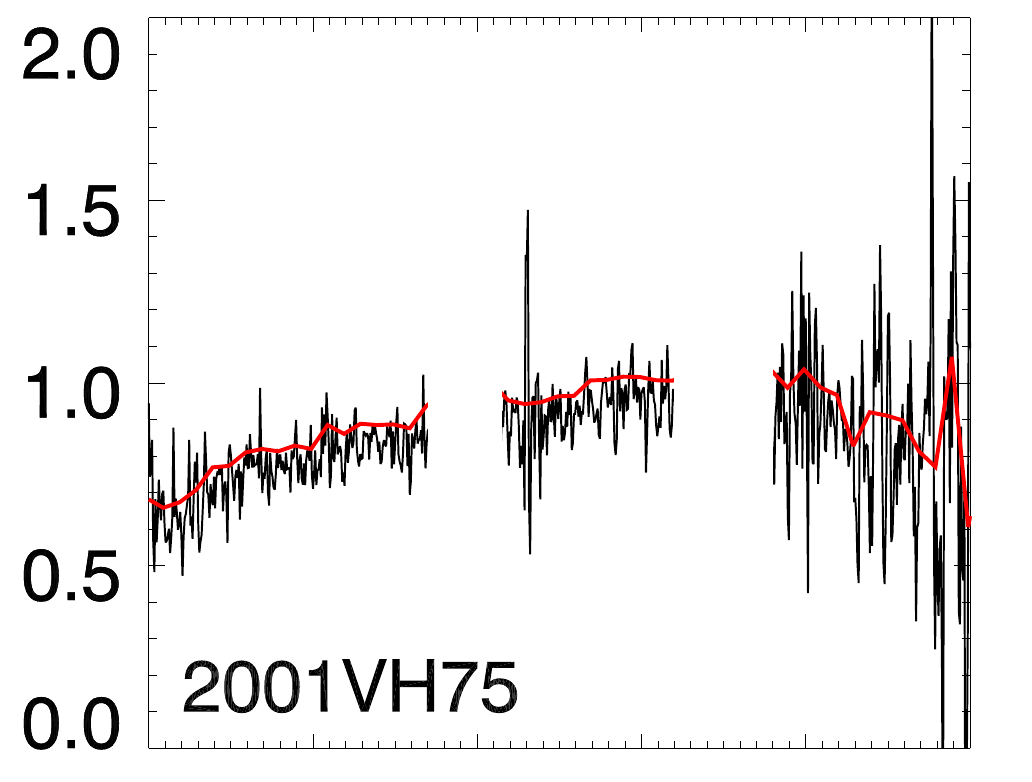}
\includegraphics[width=0.47\columnwidth,trim=0.1cm 15.0cm 9.5cm 5cm,clip=true]{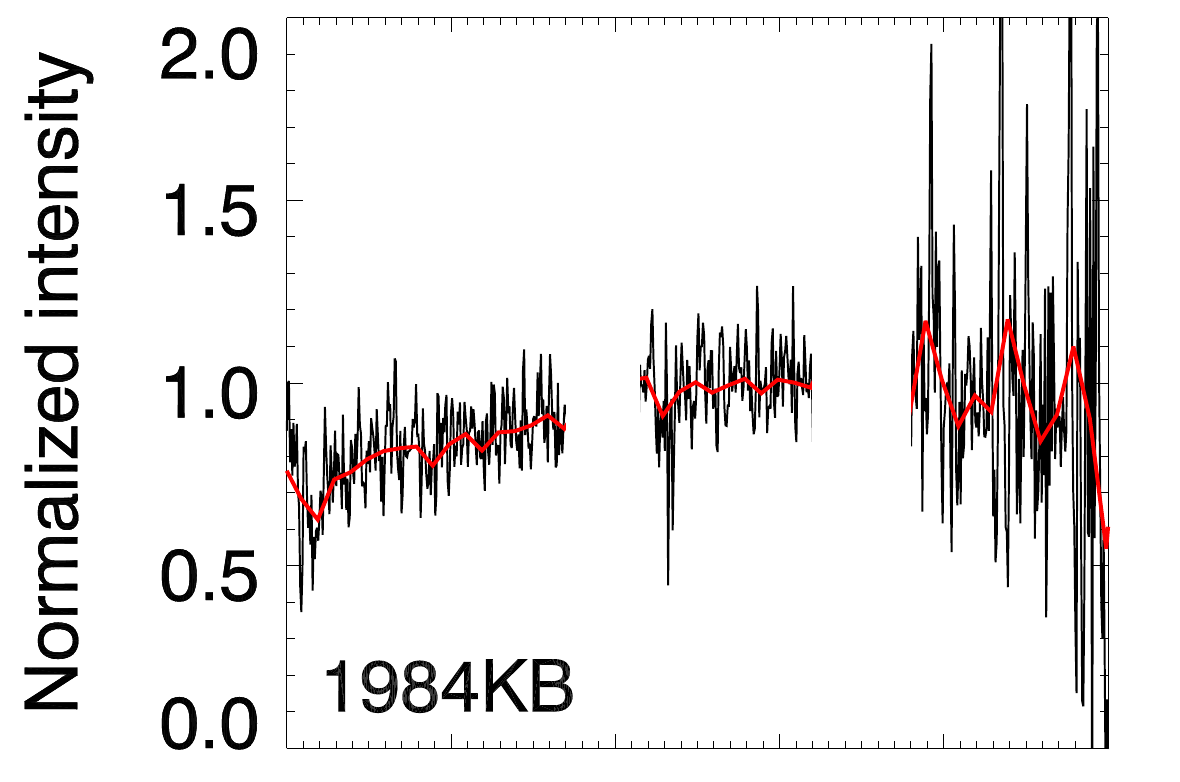}
\includegraphics[width=0.415\columnwidth,trim=1.5cm 15.0cm 9.6cm 5cm,clip=true]{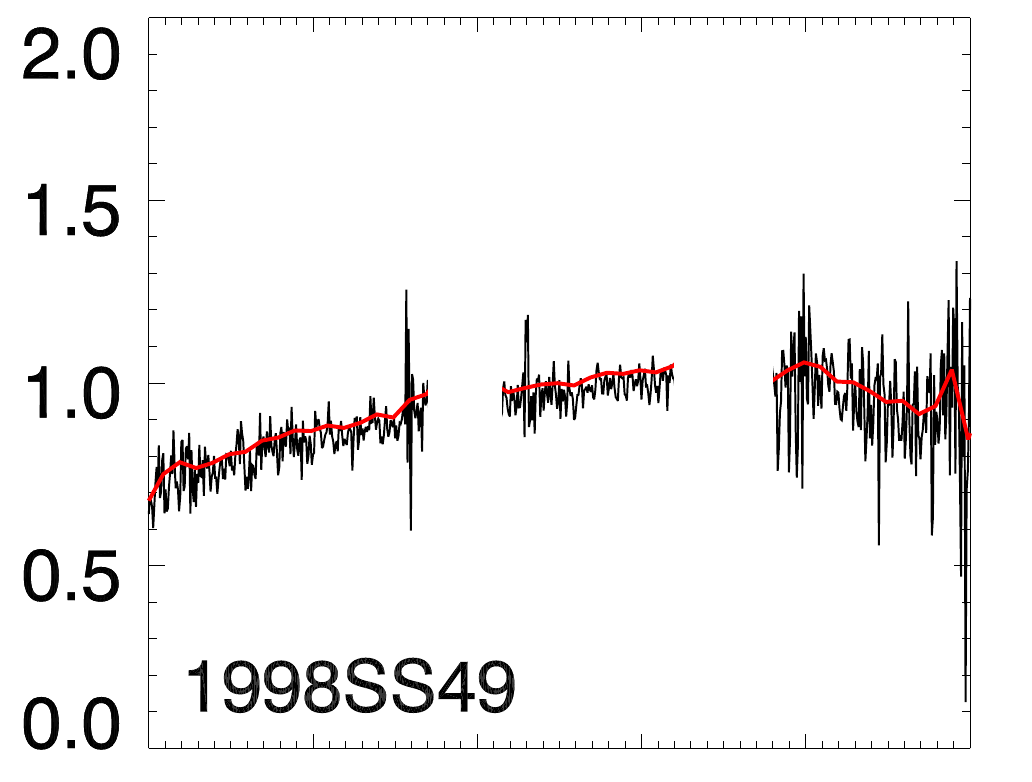}
\includegraphics[width=0.47\columnwidth,trim=0.1cm 12.5cm 9.5cm 5cm,clip=true]{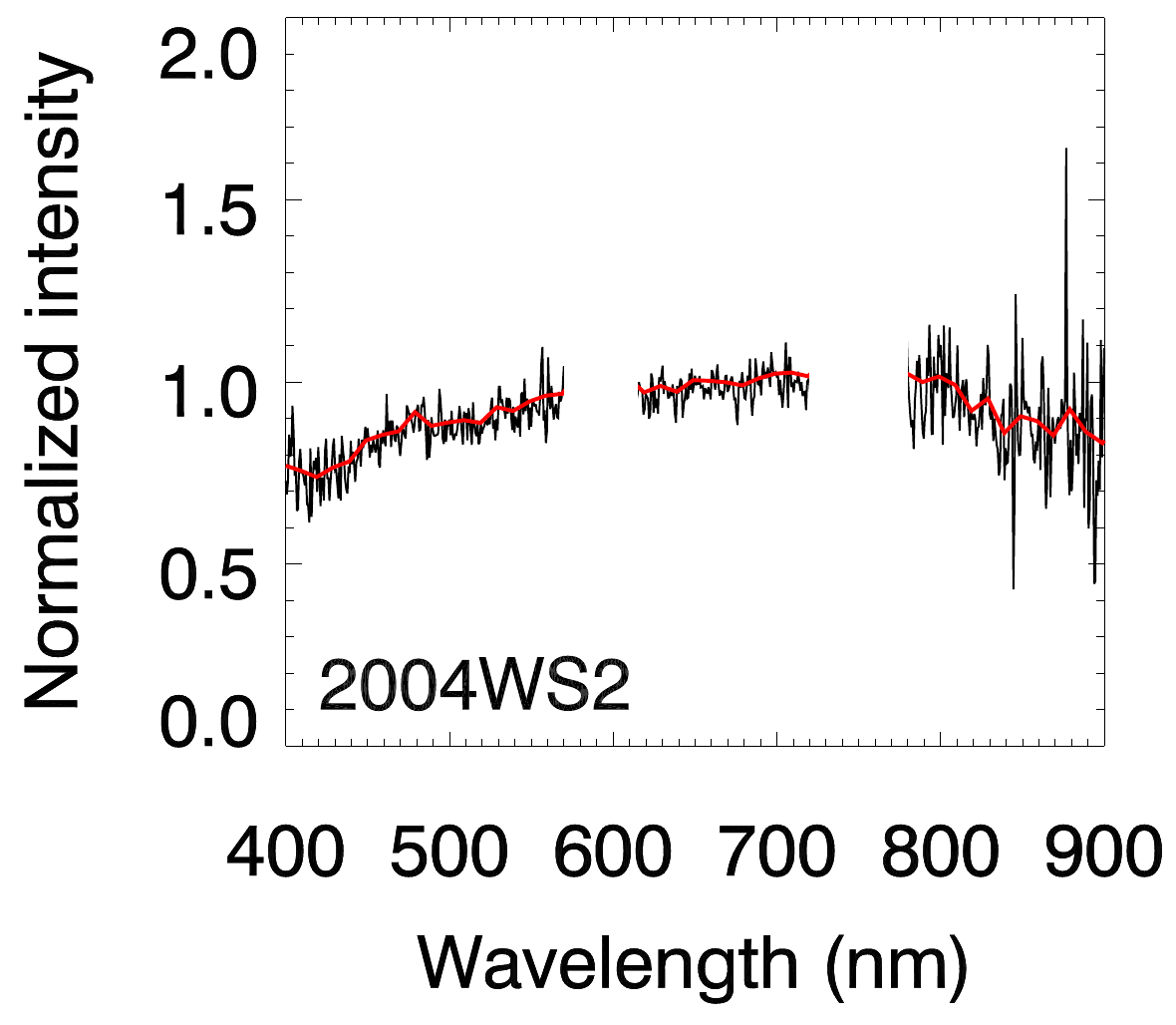}
\includegraphics[width=0.415\columnwidth,trim=1.5cm 12.5cm 9.6cm 5cm,clip=true]{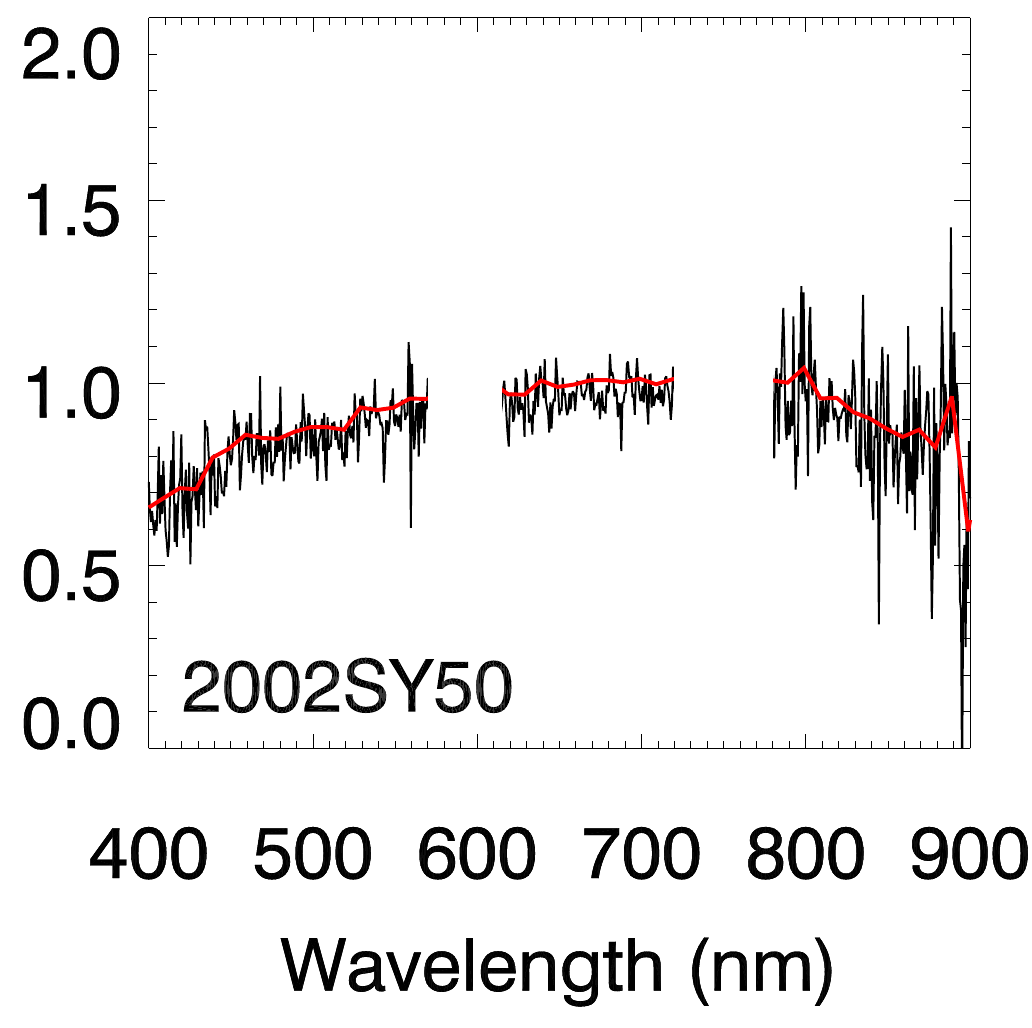}
\caption{Low resolution spectra of 1999~VT$_{25}$, 2001~VH$_{75}$, 1984~KB, 1998~SS$_{49}$, 2004~WS$_{2}$, and 2002~SY$_{50}$.}
\label{Spectra-NEOs2}
\end{figure}

Figure \ref{Spectra-NEOs2} displays the spectra of the NEOs belonging to Group 2. Also in this case the spectra do not show any absorption or emission feature for $\lambda$ < 800 nm. 1984~KB, 1998~SS$_{49}$, 2001~VH$_{75}$, 2002~SY$_{50}$, and 2004~WS$_{2}$ show the silicate absorption at the red-end of the spectrum.
While 1999~VT$_{25}$ is flat, the other objects display a moderate reddening, in the selected wavelength range, comparable with the one of 2P/Encke. 
Looking at the orbital plots (Fig. \ref{Orb-plot}), the orbit of 1999~VT$_{25}$ is quite different from the one of 2P/Encke. Even though this object was considered a possible family candidate, it can instead be an interloper.

\begin{figure}[h]
\centering
\includegraphics[width=0.47\columnwidth,trim=0.1cm 12.5cm 9.5cm 5cm,clip=true]{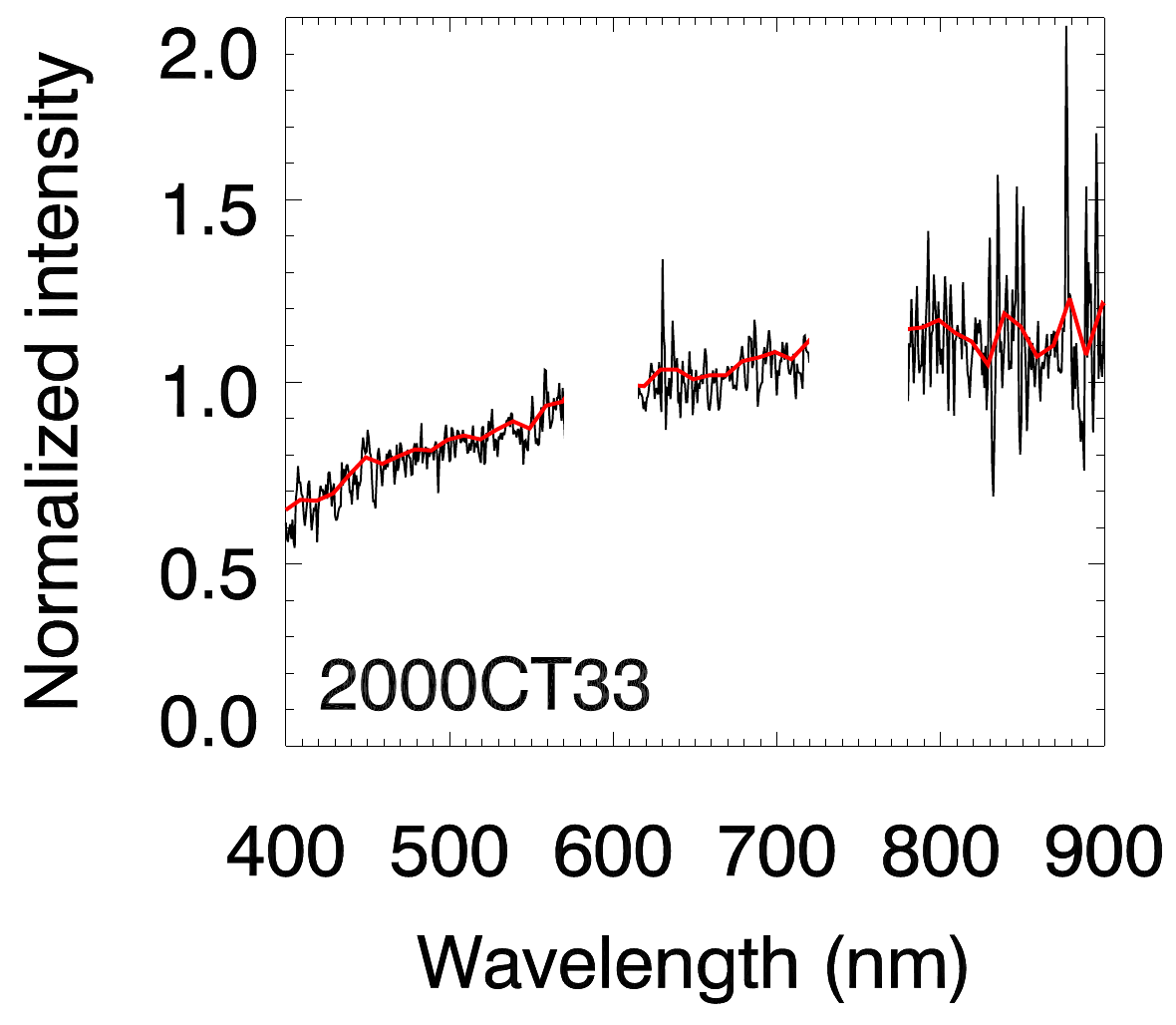}
\includegraphics[width=0.415\columnwidth,trim=1.5cm 12.5cm 9.6cm 5cm,clip=true]{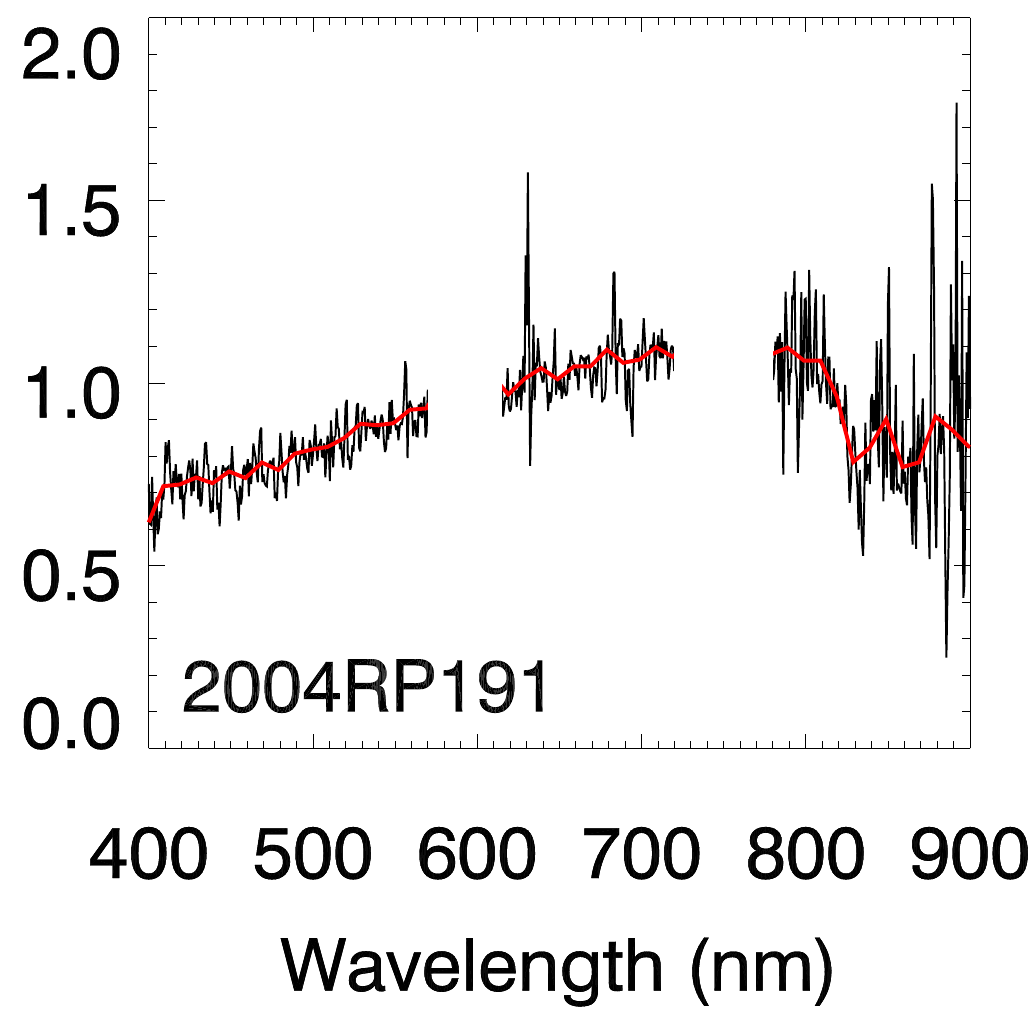}
\caption{Low resolution spectra of 2000~CT$_{33}$ and 2004~RP$_{191}$.}
\label{Spectra-NEOs3}
\end{figure}

For completeness, Fig. \ref{Spectra-NEOs3} show the spectra of 2000~CT$_{33}$ and 2004~RP$_{191}$, two unrelated asteroids observed on the same night. These spectra are featureless at $\lambda$ < 800 nm and they are redder than the one of 2P/Encke. The latter spectrum also displays a decrease in flux towards the red-end of the spectrum, typical of S-type asteroids. Since these two object are not linked with 2P/Encke, we do not expect them to show spectral similarities.

\begin{figure}[h]
\centering
\includegraphics[width=0.95\columnwidth,trim=0.5cm 11.5cm 6.5cm 4cm,clip=true]{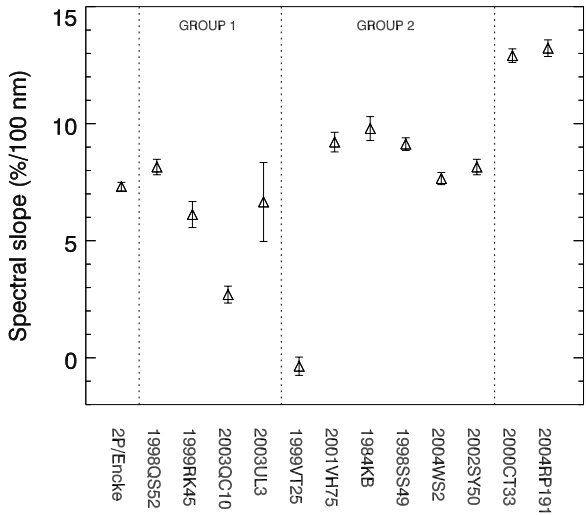}
\caption{Spectral slope of 2P/Encke and the selected NEOs in the wavelength range $\Delta \lambda$ = 420 -- 750 nm, as summarised in Tab. \ref{slope-spectra}.}
\label{spectral-slope-fig}
\end{figure}

In conclusion, the spectra of 2P/Encke and of all the selected NEOs are featureless for $\lambda$ < 800 nm. 1998~QS$_{52}$ and all NEOs belonging to Group 2 (except for 1999~VT$_{25}$) show the silicate absorption at the red-end of the spectrum. Except for 2003~QC$_{10}$ and 1999~VT$_{25}$ that have a flatter spectrum, the other objects belonging to Group 1 and Group 2 show a moderate reddening slope at $\lambda$ < 800 nm, compatible with the one of 2P/Encke, as it can be seen in Fig. \ref{spectral-slope-fig}. 

It is difficult to draw a firm conclusion if those NEOs and 2P/Encke are linked just based on their spectral properties since the only available parameter is the spectral slope. If we believe that the spectral slope per se is indicative of belonging to the same family, then 2P/Encke and 1998~QS$_{52}$, 1999~RK$_{45}$, 2003~UL$_{3}$, 2001~VH$_{75}$, 1984~KB, 1998~SS$_{49}$, 2004~WS$_{2}$, and 2002~SY$_{50}$ might be related. The absence of absorption features is a common characteristic to other comets as well, thus it cannot be used to enforce the link between 2P/Encke and the selected NEOs, but does at least demonstrate that Taurid complex NEOs are compatible with a cometary source based on spectral slopes.

However, including the information contained in the noisier part of the spectrum in the 800 > $\lambda$ > 900 nm range shows that the spectral slope gives only part of the picture. In this region many of the NEOs show evidence of a weak absorption feature, most likely due to silicates, which is not seen in cometary spectra.

\subsection{NEOs classification based on the shape of their spectrum}
\label{sec-classification}
We have classified our targets according to the Bus-DeMeo taxonomy \citep{DeMeo2009}, which uses both optical and near-IR spectra to identify 24 spectral classes existing within the asteroid population. This classification scheme is an update of the original Bus taxonomy that was based on optical spectra only \citep{Bus1999}. Although we only have optical data, \citet{DeMeo2009} show a close correspondence between this taxonomy with the previous optical-based system, therefore we have chosen the newer scheme as the most complete spectral classification system to date. For each spectrum we calculated the mean reflectance over 50 nm bins at each of the standard wavelengths in the taxonomic scheme (450 nm to 900 nm). 
We then simply calculated the $\chi^2$ value between the target spectrum and the 24 spectral classes to identify the best fit, checking by eye that the identified spectral matches looked reasonable. The resulting spectral classifications are given in Table \ref{spectraltype-spectra}. Where it was impossible to assign a unique classification to an object, all likely spectral classes are listed. In assigning the listed classifications we ignored the bin centres at 600 nm and 750 nm due to problems with the reflectance spectra in those regions, but noted that including them made no significant difference. Two of our targets have been spectrally classified previously. 1998~QS$_{52}$ was observed by \citet{Binzel2004} in their large study of the NEO population, and found to be Sq-class. Our classification of Sq/Sr is consistent with theirs. \citet{Popescu2011} also find this NEO to be Sr class based on near infrared data. Asteroid 1984 KB was classified as S-class by \citet{Bell1988} and Sq by \citet{Popescu2014}. Again, our classification of K/Sq places this object firmly within the S-complex of asteroids, and is consistent with the previous research. A recent paper by \citet{DeMeo2014} includes classifications for 1998~SS$_{49}$ and 2002~SY$_{50}$, which are also found to be S-types based on near infrared data.

\begin{table}[h]
\caption{Spectral type of the objects.}
\label{spectraltype-spectra}
\begin{center}
\begin{tabular}{lc}
\hline
Object	& Spectral type \\
\hline
2P/Encke		& Xe \\
\hline
Group 1 Taurids:	&	\\
\hline
1998~QS$_{52}$	& Sq/Sr	\\
1999~RK$_{45}$	& Q	\\
2003~QC$_{10}$	& C/Xn \\
2003~UL$_{3}$		& V/O/Q \\
\hline
Group 2 Taurids:	&	\\
\hline
1984~KB		& K/Sq \\
1998~SS$_{49}$	& Sq/Sr \\
1999~VT$_{25}$	& C \\
2001~VH$_{75}$	& Q \\
2002~SY$_{50}$	& Q \\
2004~WS$_{2}$		& Q \\
\hline
Non-Taurid asteroids:	&	\\
2000~CT$_{33}$	& L/D \\
2004~RP$_{191}$	& R/Sa \\
\hline
\end{tabular}
\end{center}
\end{table}

Going by the groups listed in Table \ref{obs-table}, 2P/Encke is identified as an X-class object in this scheme, with the best match being Xe. An X-class classification is no surprise, as the X-complex comprises of E, M and P-class asteroids, all of which show a moderate red slope with no spectral features that can only be differentiated by albedo knowledge. This exactly describes comet nuclei with moderate red spectral slopes. Three of the Group 1 Taurids are clearly silicate, although the S/N in the spectrum of 2003~UL$_{3}$ is too low for a clear
classification. 2003~QC$_{10}$ is likely a C-type asteroid; the formal fitting also gives Xn as a possible classification, but this relies on the existence of a Nysa-like absorption feature at 900 nm which is in the region of low S/N in our spectra. Similarly, all but one of the Group 2 Taurids are also silicate. We could not distinguish between the Cb/Ch/Cgh/C classifications for 1999~VT$_{25}$, so we simply adopt C. Finally, we list for completeness the spectral classes of the two non-Taurid NEOs. 

Based on the spectral classification, none of the Taurid NEOs is similar to 2P/Encke. Apart from 2003~QC$_{10}$ and 1999~VT$_{25}$ that are C-type, all the NEOs are silicate types. If we also consider the recent results from \citet{Popescu2014} we see a strong preference for S-complex asteroids in Taurid-like orbits. Two further objects from their sample are part of our `Group 1' list, with the most secure association with the Taurids: (4183) Cuno and (4486) Mithra, found to be Q and Sq type respectively. Two of their other targets [(2201) Oljato and (5143) Heracles] would fit into our Group 2, and are also Q or Sq types, while they only identified one C-type, (269690) 1996~RG$_3$, which was furthest from the centre of the Taurid cloud by their definition, and also fits into our Group 2. This suggests that these objects are not unusual for near-Earth and inner main belt asteroids, where S-types dominate, and do not require a large split comet to explain their origin.

\section{Results: 2P/Encke and CM chondrites}
The next step in the analysis was the comparison between the spectra of the nucleus of 2P/Encke and the NEOs and the ones of CM chondrites, the class to which Maribo and Sutter's Mill belong. \newline
All the spectra we have obtained are featureless below 800 nm and relatively flat (see spectral slopes in Tab. \ref{slope-spectra}). A tentative classification based on the shape of the spectra is described in Sec. \ref{sec-classification}.
The spectrum of 2P/Encke resembles one of the primitive asteroids, which are believed to be associated with carbonaceous chondrites. A link between Encke and the CM chondrites, including the Maribo and Sutter's Mill meteorites, is therefore possible. 

Although Maribo is too small to have a reflectance spectrum measured (a destructive process), a sufficient number of samples were collected of Sutter's Mill, and a spectrum covering the 0.3 - 5 $\mu$m range is available \citep[their Fig.~S8]{Jenniskens2012Sci}.
CM chondrites are carbonaceous chondrites from a parent body that was originally rich in volatiles including liquid water and organics. Aqueous alteration of metal and silicate phases resulted in the formation of phyllosilicates and iron alteration minerals. The altered phases are seen as absorption features in reflected light at 700 nm and 3 $\mu m$ \citep{Rivkin2002}. Since our observations cover the wavelength range 400 - 900 nm, we should therefore be able to observe the 700 nm, but not the 3 $\mu m$ feature. \newline 
As can be seen in Fig. \ref{Spectrum-Encke}, \ref{Spectra-NEOs1}, and \ref{Spectra-NEOs2} none of the observed objects show signs of an absorption feature at 700 nm. Therefore  our observations do not provide a direct link between any of the observed NEOs and CM chondrites. The remaining question is if the data rule out that the CM chondrites could be associated with the observed objects. 

The fact that 2P/Encke's surface spectrum does not match that of a carbonaceous chondrite does not rule out that 2P/Encke could be the parent body of Maribo and/or Sutter's Mill. It is important to remember that CM chondrites have highly variable spectral properties and that comet surfaces undergo significant changes due to Solar heating. The 700 nm features is caused by hydrated minerals which disappear if the material is moderately heated \citep{Hiroi1996}. It is therefore possible that the surface spectra of comets and primitive asteroids do not give a true representation of the interior composition, which is the volume sampled when looking at meteorite sample.


\section{Summary and conclusions}
We analysed R filter images and low resolution spectra of 2P/Encke and 10 NEOs obtained at the ESO-VLT using FORS on August 2, 2011.
\begin{enumerate}
\item The comparison of the surface brightness profile of the objects with the one of a star indicates that no detectable activity is present. 
\item The spectra of 2P/Encke and of the NEOs are featureless for $\lambda$ < 800 nm. 1998~QS$_{52}$ and all NEOs belonging to Group 2 (except for 1999~VT$_{25}$) show the silicate absorption at the red end of the spectrum.
With the exception of 2003~QC$_{10}$ and 1999~VT$_{25}$ that show flatter spectra, the other objects belonging to Group 1 and 2 show a moderate reddening slope ($\sim$ 6-10 \%/100 nm), compatible with the one of 2P/Encke. It is however difficult to draw a firm conclusion if 2P/Encke and the selected NEOs are linked just based on their spectral properties, since the only available parameter is the spectral slope. The absence of absorption features at $\lambda$ < 800 nm is a common characteristic to other comets as well, thus cannot be used to enforce the link between 2P/Encke and the observed NEOs.
\item The absence of the 700 nm absorption feature in the spectra of 2P/Encke and the selected NEOs doesn't allow us to establish a direct link between 2P/Encke, the observed NEOs and CM chondrites. 
\item The spectral classification of the objects shows that none of the Taurid NEOs is similar to 2P/Encke when the low S/N range ($\lambda$ > 800 nm) is taken into account. The best fit taxonomic group for most of the NEOs is the S-complex, while Encke doesn't show any evidence for absorption at $\lambda$ > 800 nm, and is best fit by the Xe-type template.
\item The difference between the spectrum of 2P/Encke and CM chondrites are due to the presence of hydrated minerals in CM chondrites. A connection between 2P/Encke remains ambiguous as we do not know if 2P/Encke's surface has been dehydrated due Solar heating.  
\end{enumerate}
We conclude that there is no evidence that NEOs with Taurid-complex orbits exhibit any evidence of a compositional association with 2P/Encke, and that the presence of potentially unrelated bodies in these orbits means that it is still impossible to choose between a cometary or asteroidal source for the Taurid CM chondrite meteorites. If, instead, we assume that the whole Taurid-complex does have a single parent body, this (presumably large) object must have contained significant inhomogeneity, to produce some cometary and some S-type fragments.

\begin{acknowledgements}
  
CS received funding from the European Union Seventh Framework Programme (FP7/2007-2013) under grant agreement no. 268421. HH received funding from the Danish National Research Foundation. We thank the anonymous referee for helpful comments that improved this paper.

  \end{acknowledgements}

\bibliographystyle{aa} 
\bibliography{cecibib}  

\begin{thebibliography}{40}
\expandafter\ifx\csname natexlab\endcsname\relax\def\natexlab#1{#1}\fi

\bibitem[{{Appenzeller} {et~al.}(1998){Appenzeller}, {Fricke}, {F{\"u}rtig},
  {G{\"a}ssler}, {H{\"a}fner}, {Harke}, {Hess}, {Hummel}, {J{\"u}rgens},
  {Kudritzki}, {Mantel}, {Meisl}, {Muschielok}, {Nicklas}, {Rupprecht},
  {Seifert}, {Stahl}, {Szeifert}, \& {Tarantik}}]{Appenzeller1998}
{Appenzeller}, I., {Fricke}, K., {F{\"u}rtig}, W., {et~al.} 1998, The
  Messenger, 94, 1

\bibitem[{{Asher} {et~al.}(1993){Asher}, {Clube}, \& {Steel}}]{Asher1993MNRAS}
{Asher}, D.~J., {Clube}, S.~V.~M., \& {Steel}, D.~I. 1993, \mnras, 264, 93

\bibitem[{{Bell} {et~al.}(1988){Bell}, {Hawke}, \& {Brown}}]{Bell1988}
{Bell}, J.~F., {Hawke}, B.~R., \& {Brown}, R.~H. 1988, \icarus, 73, 482

\bibitem[{{Binzel} {et~al.}(2004){Binzel}, {Rivkin}, {Stuart}, {Harris}, {Bus},
  \& {Burbine}}]{Binzel2004}
{Binzel}, R.~P., {Rivkin}, A.~S., {Stuart}, J.~S., {et~al.} 2004, \icarus, 170,
  259

\bibitem[{{B\"ohnhardt} {et~al.}(2008){B\"ohnhardt}, {Tozzi}, {Bagnulo},
  {Muinonen}, {Nathues}, \& {Kolokolova}}]{Boehnhardt2008aa}
{B\"ohnhardt}, H., {Tozzi}, G.~P., {Bagnulo}, S., {et~al.} 2008, \aap, 489,
  1337

\bibitem[{{Bus}(1999)}]{Bus1999}
{Bus}, S.~J. 1999, PhD thesis, MASSACHUSETTS INSTITUTE OF TECHNOLOGY

\bibitem[{{Clube} \& {Napier}(1984)}]{Clube1984MNRAS}
{Clube}, S.~V.~M. \& {Napier}, W.~M. 1984, \mnras, 211, 953

\bibitem[{{DeMeo} {et~al.}(2014){DeMeo}, {Binzel}, \& {Lockhart}}]{DeMeo2014}
{DeMeo}, F.~E., {Binzel}, R.~P., \& {Lockhart}, M. 2014, \icarus, 227, 112

\bibitem[{{DeMeo} {et~al.}(2009){DeMeo}, {Binzel}, {Slivan}, \&
  {Bus}}]{DeMeo2009}
{DeMeo}, F.~E., {Binzel}, R.~P., {Slivan}, S.~M., \& {Bus}, S.~J. 2009,
  \icarus, 202, 160

\bibitem[{{Fern{\'a}ndez} {et~al.}(2000){Fern{\'a}ndez}, {Lisse}, {K{\"a}ufl},
  {Peschke}, {Weaver}, {A'Hearn}, {Lamy}, {Livengood}, \&
  {Kostiuk}}]{Fernandez2000ic}
{Fern{\'a}ndez}, Y.~R., {Lisse}, C.~M., {K{\"a}ufl}, H.~U., {et~al.} 2000,
  Icarus, 147, 145

\bibitem[{{Fern{\'a}ndez} {et~al.}(2005){Fern{\'a}ndez}, {Lowry}, {Weissman},
  {Mueller}, {Samarasinha}, {Belton}, \& {Meech}}]{Fernandez2005}
{Fern{\'a}ndez}, Y.~R., {Lowry}, S.~C., {Weissman}, P.~R., {et~al.} 2005,
  \icarus, 175, 194

\bibitem[{{Haack} {et~al.}(2012){Haack}, {Grau}, {Bischoff}, {Horstmann},
  {Wasson}, {S{\o}rensen}, {Laubenstein}, {Ott}, {Palme}, {Gellissen},
  {Greenwood}, {Pearson}, {Franchi}, {Gabelica}, \&
  {Schmitt-Kopplin}}]{Haack2012M}
{Haack}, H., {Grau}, T., {Bischoff}, A., {et~al.} 2012, Meteoritics and
  Planetary Science, 47, 30

\bibitem[{{Haack} {et~al.}(2011){Haack}, {Michelsen}, {Stober}, {Keuer},
  {Singer}, \& {Williams}}]{Haack2011}
{Haack}, H., {Michelsen}, R., {Stober}, G., {et~al.} 2011, Meteoritics and
  Planetary Science Supplement, 74, 5271

\bibitem[{{Hainaut} {et~al.}(2012){Hainaut}, {Boehnhardt}, \&
  {Protopapa}}]{Hainaut2012}
{Hainaut}, O.~R., {Boehnhardt}, H., \& {Protopapa}, S. 2012, \aap, 546, A115

\bibitem[{{Harmon} \& {Nolan}(2005)}]{HarmonNolan2005}
{Harmon}, J.~K. \& {Nolan}, M.~C. 2005, \icarus, 176, 175

\bibitem[{{Hiroi} {et~al.}(1996){Hiroi}, {Zolensky}, {Pieters}, \&
  {Lipschutz}}]{Hiroi1996}
{Hiroi}, T., {Zolensky}, M.~E., {Pieters}, C.~M., \& {Lipschutz}, M.~E. 1996,
  Meteoritics and Planetary Science, 31, 321

\bibitem[{{Hsieh} \& {Jewitt}(2006)}]{Hsieh2006}
{Hsieh}, H.~H. \& {Jewitt}, D. 2006, Science, 312, 561

\bibitem[{{Jenniskens} {et~al.}(2012{\natexlab{a}}){Jenniskens}, {Fries},
  {Yin}, {Zolensky}, {Krot}, {Sandford}, {Sears}, {Beauford}, {Ebel},
  {Friedrich}, {Nagashima}, {Wimpenny}, {Yamakawa}, {Nishiizumi}, {Hamajima},
  {Caffee}, {Welten}, {Laubenstein}, {Davis}, {Simon}, {Heck}, {Young}, {Kohl},
  {Thiemens}, {Nunn}, {Mikouchi}, {Hagiya}, {Ohsumi}, {Cahill}, {Lawton},
  {Barnes}, {Steele}, {Rochette}, {Verosub}, {Gattacceca}, {Cooper}, {Glavin},
  {Burton}, {Dworkin}, {Elsila}, {Pizzarello}, {Ogliore}, {Schmitt-Kopplin},
  {Harir}, {Hertkorn}, {Verchovsky}, {Grady}, {Nagao}, {Okazaki}, {Takechi},
  {Hiroi}, {Smith}, {Silber}, {Brown}, {Albers}, {Klotz}, {Hankey}, {Matson},
  {Fries}, {Walker}, {Puchtel}, {Lee}, {Erdman}, {Eppich}, {Roeske},
  {Gabelica}, {Lerche}, {Nuevo}, {Girten}, \& {Worden}}]{Jenniskens2012Sci}
{Jenniskens}, P., {Fries}, M.~D., {Yin}, Q.-Z., {et~al.} 2012{\natexlab{a}},
  Science, 338, 1583

\bibitem[{{Jenniskens} {et~al.}(2012{\natexlab{b}}){Jenniskens}, {Girten},
  {Sears}, {Sandford}, {Cooper}, {Ehrgott}, {Koop}, {Albers}, {Fries}, {Klotz},
  {Hankey}, {Schmidt}, \& {Worden}}]{Jenniskens2012}
{Jenniskens}, P., {Girten}, B., {Sears}, D., {et~al.} 2012{\natexlab{b}},
  Meteoritics and Planetary Science Supplement, 75, 5376

\bibitem[{{Jewitt}(2012)}]{Jewitt2012AJ}
{Jewitt}, D. 2012, \aj, 143, 66

\bibitem[{{Jewitt}(2002)}]{Jewitt2002}
{Jewitt}, D.~C. 2002, \aj, 123, 1039

\bibitem[{{Jopek} \& {Williams}(2013)}]{Jopek2013MNRAS}
{Jopek}, T.~J. \& {Williams}, I.~P. 2013, \mnras, 430, 2377

\bibitem[{{Kelley} {et~al.}(2013){Kelley}, {Fern{\'a}ndez}, {Licandro},
  {Lisse}, {Reach}, {A'Hearn}, {Bauer}, {Campins}, {Fitzsimmons}, {Groussin},
  {Lamy}, {Lowry}, {Meech}, {Pittichov{\'a}}, {Snodgrass}, {Toth}, \&
  {Weaver}}]{Kelley2013}
{Kelley}, M.~S., {Fern{\'a}ndez}, Y.~R., {Licandro}, J., {et~al.} 2013,
  \icarus, 225, 475

\bibitem[{{Lamy} \& {Toth}(2009)}]{Lamy+Toth09}
{Lamy}, P. \& {Toth}, I. 2009, Icarus, 201, 674

\bibitem[{{Lamy} {et~al.}(2004){Lamy}, {Toth}, {Fernandez}, \&
  {Weaver}}]{Lamy2004comets}
{Lamy}, P.~L., {Toth}, I., {Fernandez}, Y.~R., \& {Weaver}, H.~A. 2004, {The
  sizes, shapes, albedos, and colors of cometary nuclei} (University of Arizona
  Press, Tucson), 223--264

\bibitem[{{Levison} {et~al.}(2006){Levison}, {Terrell}, {Wiegert}, {Dones}, \&
  {Duncan}}]{Levison2006}
{Levison}, H.~F., {Terrell}, D., {Wiegert}, P.~A., {Dones}, L., \& {Duncan},
  M.~J. 2006, \icarus, 182, 161

\bibitem[{{Lodders} \& {Osborne}(1999)}]{Lodders1999SSRv}
{Lodders}, K. \& {Osborne}, R. 1999, \ssr, 90, 289

\bibitem[{{Luu} \& {Jewitt}(1990)}]{Luu1990}
{Luu}, J. \& {Jewitt}, D. 1990, \icarus, 86, 69

\bibitem[{{Luu}(1993)}]{Luu1993ic}
{Luu}, J.~X. 1993, Icarus, 104, 138

\bibitem[{{Meeus}(1998)}]{Meeus1998}
{Meeus}, J. 1998, {Astronomical algorithms} (Willmann-Bell Inc.)

\bibitem[{{Michelsen} {et~al.}(2006){Michelsen}, {Nathues}, \&
  {Lagerkvist}}]{Michelsen2006}
{Michelsen}, R., {Nathues}, A., \& {Lagerkvist}, C.-I. 2006, \aap, 451, 331

\bibitem[{{Popescu} {et~al.}(2011){Popescu}, {Birlan}, {Binzel}, {Vernazza},
  {Barucci}, {Nedelcu}, {DeMeo}, \& {Fulchignoni}}]{Popescu2011}
{Popescu}, M., {Birlan}, M., {Binzel}, R., {et~al.} 2011, \aap, 535, A15

\bibitem[{{Popescu} {et~al.}(2014){Popescu}, {Birlan}, {Nedelcu}, {Vaubaillon},
  \& {Cristescu}}]{Popescu2014}
{Popescu}, M., {Birlan}, M., {Nedelcu}, D.~A., {Vaubaillon}, J., \&
  {Cristescu}, C.~P. 2014, \aap, 572, A106

\bibitem[{{Porub{\v c}an} {et~al.}(2006){Porub{\v c}an}, {Korno{\v s}}, \&
  {Williams}}]{Porubcan2006}
{Porub{\v c}an}, V., {Korno{\v s}}, L., \& {Williams}, I.~P. 2006,
  Contributions of the Astronomical Observatory Skalnate Pleso, 36, 103

\bibitem[{{Porub{\v c}an} {et~al.}(2004){Porub{\v c}an}, {Williams}, \&
  {Korno{\v s}}}]{Porubcan2004}
{Porub{\v c}an}, V., {Williams}, I.~P., \& {Korno{\v s}}, L. 2004, Earth Moon
  and Planets, 95, 697

\bibitem[{{Rivkin} {et~al.}(2002){Rivkin}, {Howell}, {Vilas}, \&
  {Lebofsky}}]{Rivkin2002}
{Rivkin}, A.~S., {Howell}, E.~S., {Vilas}, F., \& {Lebofsky}, L.~A. 2002,
  Asteroids III, 235

\bibitem[{{Snodgrass} {et~al.}(2006){Snodgrass}, {Lowry}, \&
  {Fitzsimmons}}]{Snodgrass2006mnras}
{Snodgrass}, C., {Lowry}, S.~C., \& {Fitzsimmons}, A. 2006, \mnras, 373, 1590

\bibitem[{{Tubiana} {et~al.}(2008){Tubiana}, {Barrera}, {Drahus}, \&
  {B\"ohnhardt}}]{Tubiana2008aa}
{Tubiana}, C., {Barrera}, L., {Drahus}, M., \& {B\"ohnhardt}, H. 2008, \aap,
  490, 377

\bibitem[{{Tubiana} {et~al.}(2011){Tubiana}, {B{\"o}hnhardt}, {Agarwal},
  {Drahus}, {Barrera}, \& {Ortiz}}]{Tubiana2011}
{Tubiana}, C., {B{\"o}hnhardt}, H., {Agarwal}, J., {et~al.} 2011, \aap, 527,
  A113

\bibitem[{{Whipple}(1940)}]{Whipple1940}
{Whipple}, F.~L. 1940, The Scientific Monthly, 51, 579

\end{thebibliography}

\end{document}